\documentclass[12pt]{article}
\usepackage{graphicx}

\begin{document}

\title{Mode Competition in Gas and Semiconductor Lasers}

\author{Philip F. Bagwell \\
Purdue University, School of Electrical Engineering \\ 
West Lafayette, Indiana 47907}

\date{\today}
\maketitle

\begin{abstract}

The output spectrum of both gas and semiconductor lasers usually 
contains more than one frequency. 
Multimode operation in gas versus semiconductor lasers arises 
from different physics. In gas lasers, 
slow equilibration of the electron populations at different
energies makes each frequency an independent single-mode laser. 
The slow electron diffusion 
in semiconductor lasers, combined with the spatially varying 
optical intensity patterns of the modes, 
makes each region of space an independent single-mode laser. 
We develop a rate equation 
model for the photon number in each mode which captures all these effects. 
Plotting the photon number versus pumping rate for the competing modes, 
in both subthreshold 
and above threshold operation, illustrates the changes 
in the laser output spectrum due to 
either slow equilibration or slow diffusion of electrons.

\end{abstract}


\section{Introduction}
\indent

The basic physics of mode competition in gas and semiconductor lasers has
been known since the 1960's~\cite{ross69}-~\cite{svelto}. 
When the laser medium is in quasi-equilibrium and spatial variations
in the mode patterns can be neglected, only a single frequency appears
in the laser output. A rate equation model developed by 
Siegman~\cite{siegman71,siegman86}
describes the subthreshold, transition,
and lasing regions of operation for a single mode laser. Quantum theories
of a single mode laser~\cite{loudon} give similar (if not idential) results for the
laser output power versus pumping rate as the simpler semiclassical models.

\begin{figure}[htb]
\includegraphics[width=5in]{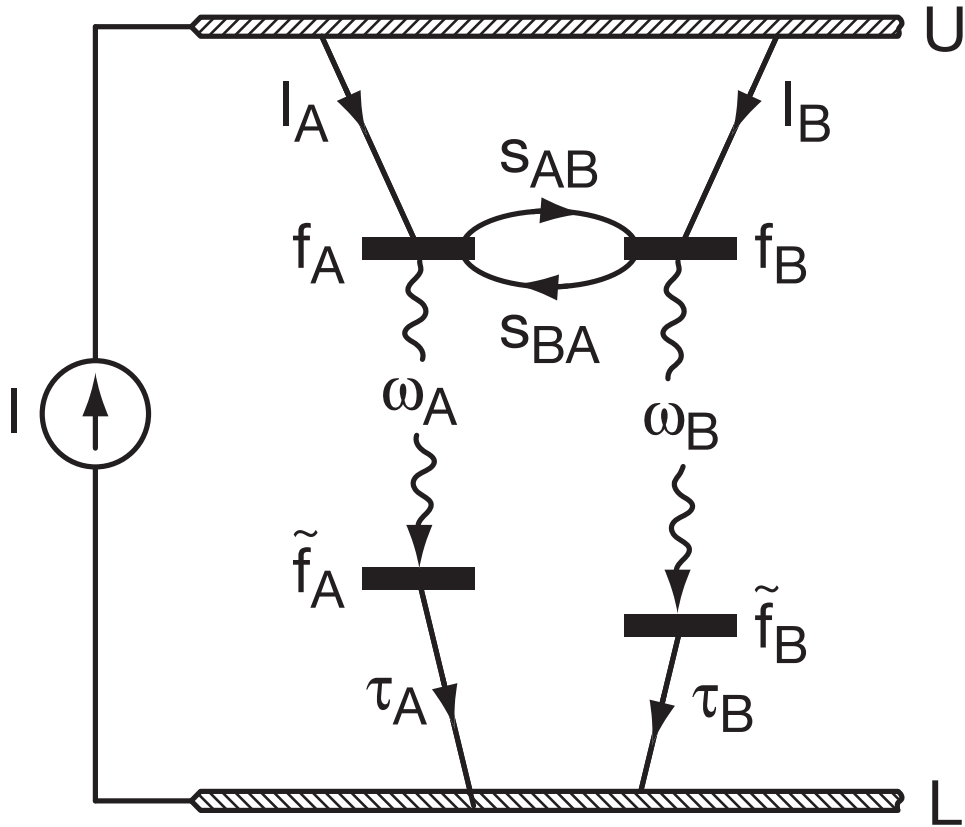}
\caption{A four level lasing system representing both gas and semiconductor
lasers. A current $I$ pumps electrons from a lower thermodynamic reservoir ($L$)
into an upper electron bath ($U$). Currents $I_A$ ($I_B$) then transfer 
electrons into the upper
level of transition $A$ ($B$). An electron
electron scattering rate $s_{AB}$ ($s_{BA}$) from the upper levels of transition
$A$ to transition $B$ ($B$ to $A$) is also present in this model.
The upper lasing levels has occupation factor
$f_A$ ($f_B$), while the occupation factor of the lower lasing level is assumed
to be $\tilde{f}_A=0$ ($\tilde{f}_B=0$).} 
\label{levels}
\end{figure}

\clearpage

In this paper we extend the rate equation models for a single model 
laser~\cite{siegman71,siegman86} to include mode competition in both gas
and semiconductor lasers. 
For the gas laser we develop a set of coupled
rate equations which allow electron scattering between the lasing levels.
The slow electron scattering rate in gas lasers allows the electron occupation
factors in the different lasing levels to get out of equilibrium with each 
other, producing multiple frequencies in the laser output.
In semiconductor lasers electron scattering between the different energy
levels is rapid, keeping the occupation factors in the different lasing
levels in equilibrium with each other. But spatial variation in the 
optical mode intensities inside the laser cavity favor
different lasing modes in different regions of space within the semiconductor.
Following Ref.~\cite{svelto} we extend the rate equations for a
homogenous (semiconductor) line to allow for spatially varying mode patterns, 
generating multiple frequencies in the laser output.

We assume both gas and semiconductor lasers are a four level system,
with two intermediate lasing levels, as shown in Fig.~\ref{levels}.
For simplicity we consider only two lasing modes, mode $A$ and mode $B$,
with mode $A$ the favored lasing mode. For the gas laser $A$ and $B$
represent localized atomic states, while for the semiconductor $A$ and $B$
are spatially extended states within the energy band of a quantum well.
To simplify the mathematics 
we assume lower lasing level is always empty, having occupation factors
$\tilde{f_A}= \tilde{f_B}=0$. 
In the language of gas lasers this means we assume
electrons in the lower lasing level empty very efficiently into an electron 
bath ($\tau_A  \; , \tau_B \to 0$). 
In semiconductor language we would say there is extremely efficient hole
capture into the quantum well. We therefore 
consider only pumping electrons into the upper lasing level. In physical
systems would also have to guarentee
charge neutrality while pumping
the laser, and therefore also consider details of pumping and
relaxation out of the lower lasing level.

In this paper we limit consideration of laser mode competition
to only 1-2 competing modes. In actual lasers many modes compete, and
this case has been considered by Casperson~\cite{kasperson}. The rate equation
models in this paper can be easily generalized to consider multiple lasing modes.

\section{Gas Lasers: Spectral Hole Burning}
\indent

\label{gaslaser}

We construct the rate equations for an inhomogeneous line following 
the example of a single mode laser from Refs.~\cite{siegman71,siegman86}. 
We consider a scattering rate $s_{A \to B} = s_{AB}$ for electrons from 
mode $A$ to mode $B$. If we firstly neglect optical transitions, the
rate equation for the occupation probability $f_A$ for electrons in state 
$A$ is
\begin{equation}
\frac{df_A}{dt} = -f_A s_{AB} (1-f_B) + f_B s_{BA} (1-f_A) + I_A.
\label{fa}
\end{equation}
Here $I_A$ is the pumping current per state.
Specializing to thermodynamic equilibrium (no pumping current) implies the
rates $s_{AB}$ and $s_{BA}$ are related by a Boltzmann factor as
\begin{equation}
s_{AB} = s_{BA} \exp{(E_A - E_B)/k_B T} . 
\end{equation}
The rate equation for the photon number $n_A$ in mode A is unchanged
from that for a single mode laser
\begin{equation}
\frac{dn_A}{dt} = K_A (n_A+1) N_A - \gamma_A  n_A.
\label{na}
\end{equation}
Here $K_A$ is an optical rate constant for the $A$ transition, $N_A$ the
number of $A$ states (number of atoms of type $A$ in a gas laser), and
$\gamma_A$ the cavity escape rate for photons having frequency $\omega_A$. 
Putting Eqs.~(\ref{fa})-(\ref{na}) together (along with analogous equations
for mode $B$) into a single matrix equation for the variables 
$f_A$, $f_B$, $n_A$, and $n_B$ gives
\begin{equation}
\frac{d}{dt}
\left[
\begin{array}{c}
f_A \cr \\
f_B \cr \\
n_A \cr \\
n_B \cr 
\end{array}
\right]
=
\left[
\begin{array}{cccc}
\begin{array}{c} 
-(K_A n_A + A_A + s_{AB}) \\
+ (s_{AB}-s_{BA})f_B 
\end{array} & s_{BA} & 0 & 0 \\
s_{AB} & 
\begin{array}{c} 
-(K_B n_B + A_B + s_{BA}) \\ 
+ (s_{BA}-s_{AB})f_A 
\end{array} & 0 & 0 \\ \\
K_A (n_A +1) N_A & 0 & -\gamma_A & 0 \cr \\
0 & K_B (n_B +1) N_B & 0 & -\gamma_B \cr \\
\end{array}
\right]
\left[
\begin{array}{c}
f_A \cr \\ 
f_B \cr \\
n_A \cr \\
n_B \cr 
\end{array}
\right]
+
\left[
\begin{array}{c}
I_A \cr \\
I_B \cr \\
0 \cr \\
0 \cr 
\end{array}
\right].
\label{spectralmc}
\end{equation}

We now specialize to $(E_A-E_B) \ll k_B T$ so that we can
approximate $s_{AB} \simeq s_{BA} = s$. 
As scattering rate $s$ becomes large ($s \to \infty$), the occupation
factors $f_A$ and $f_B$ are forced to equal each other $(f_A = f_B)$ 
in this approximation. Without the approximation $s_{AB} \simeq s_{BA}$,
and in the absence of any optical transitions, we would have the 
occupation factors forced towards a Fermi distribution $f(E)$
having $f_A = f(E_A)$ and $f_B = f(E_B)$ as $s_{AB} \to \infty$. 
The approximation $s_{AB} \simeq s_{BA}$ therefore makes only a minor correction to 
the occupation factors, and is therefore not essential for our analysis 
of mode competition. We therefore approximate $s_{AB} \simeq s_{BA} = s$ leading to
\begin{equation}
\frac{d}{dt}
\left[
\begin{array}{c}
f_A \\
f_B \\
n_A \\
n_B
\end{array}
\right]
=
\left[
\begin{array}{cccc}
-(K_A n_A + A_A + s) & s & 0 & 0 \\
s & -(K_B n_B + A_B + s) & 0 & 0 \\
K_A (n_A +1) N_A & 0 & -\gamma_A & 0 \\
0 & K_B (n_B +1) N_B & 0 & -\gamma_B
\end{array}
\right]
\left[
\begin{array}{c}
f_A \\
f_B \\
n_A \\
n_B
\end{array}
\right]
+
\left[
\begin{array}{c}
I_A \\
I_B \\
0 \\
0
\end{array}
\right].
\label{sspectralmc}
\end{equation}

Inspection of the upper left quadrant of the matrix in Eq.~(\ref{sspectralmc}) 
shows that the
scattering rate $s$ is negligible until $s$ exceeds one of the spontaneous 
emission rates $A_A$ or $A_B$.
Thus, the transition from two independent lasers ($s=0$) to a homogeneous
line ($s \to \infty$) occurs when the scattering rate $s$ exceeds the spontaneous 
emission rates $A_A$ and $A_B$.
This is true in open cavity lasers with luminescence through the 
sides of the laser cavity,
so that the spontaneous emission rates $A_A$ and $A_B$ greatly exceed 
the cavity escape rates
$\gamma_A$ and $\gamma_B$. If the cavity is closed, so that no side 
luminescence occurs, the
spontaneous emission rates are forced towards the cavity rates, 
i.e. $A_A \to \gamma_A$ and 
$A_B \to \gamma_B$.  In the case of a closed cavity the transition 
from two independent
lasers to a homogeneous line
occurs when the scattering rate $s$ is comparable to the 
cavity rates $\gamma_A$ and $\gamma_B$. 
Note that on a homogeneous line $(s \to \infty)$ Eq.~(\ref{sspectralmc})
simplifies to $f \equiv f_A = f_B$ and
\begin{equation}
\frac{d}{dt}
\left[
\begin{array}{c}
2f \\
n_A \\
n_B
\end{array}
\right]
=
\left[
\begin{array}{cccc}
-(K_A n_A + K_B n_B + A) & 0 & 0 \\
K_A (n_A +1) N_A & -\gamma_A & 0 \\
K_B (n_B +1) N_B & 0 & -\gamma_B
\end{array}
\right]
\left[
\begin{array}{c}
f \\
n_A \\
n_B
\end{array}
\right]
+
\left[
\begin{array}{c}
I \\
0 \\
0
\end{array}
\right],
\label{homogline}
\end{equation}
where $I = I_A + I_B$ and $A = A_A + A_B$.

We solve Eq.~(\ref{sspectralmc}) in steady state using an iterative technique.
From the $i$th iteration for the variables $f_A^i$, $f_B^i$, $n_A^i$, and $n_B^i$,
we produce the $(i+1)$st iteration by
\begin{equation}
-\left[
\begin{array}{c}
f_A^{i+1} \\
f_B^{i+1}  \\
n_A^{i+1}  \\
n_B^{i+1} 
\end{array}
\right]
=
\left[
\begin{array}{cccc}
-(K_A n_A^{i}  + A_A + s) & s & 0 & 0 \\
s & -(K_B n_B^{i} + A_B + s) & 0 & 0 \\
K_A (n_A^{i} +1) N_A & 0 & -\gamma_A & 0 \\
0 & K_B (n_B^{i}+1) N_B & 0 & -\gamma_B
\end{array}
\right]^{-1}
\left[
\begin{array}{c}
I_A \\
I_B \\
0 \\
0
\end{array}
\right].
\label{iterate}
\end{equation}
We have tried several different types of initial guesses for
$f_A$, $f_B$, $n_A$, and $n_B$ to start the iterative
procedure, and the final results seem to be independent of the 
different initial guesses. The initial guess which seems to converge
in the shortest time is to start in the subthreshold region and
take for $f_A^i$, $f_B^i$, $n_A^i$, 
and $n_B^i$ the analytical results for two independent single mode lasers ($s=0$).
When incrementing to the next pumping rate, assume an initial guess
for $f_A$, $f_B$, $n_A$, and $n_B$ which are just the converged values
at the previous pumping rate.

\begin{figure}[htb]
\includegraphics[width=3in]{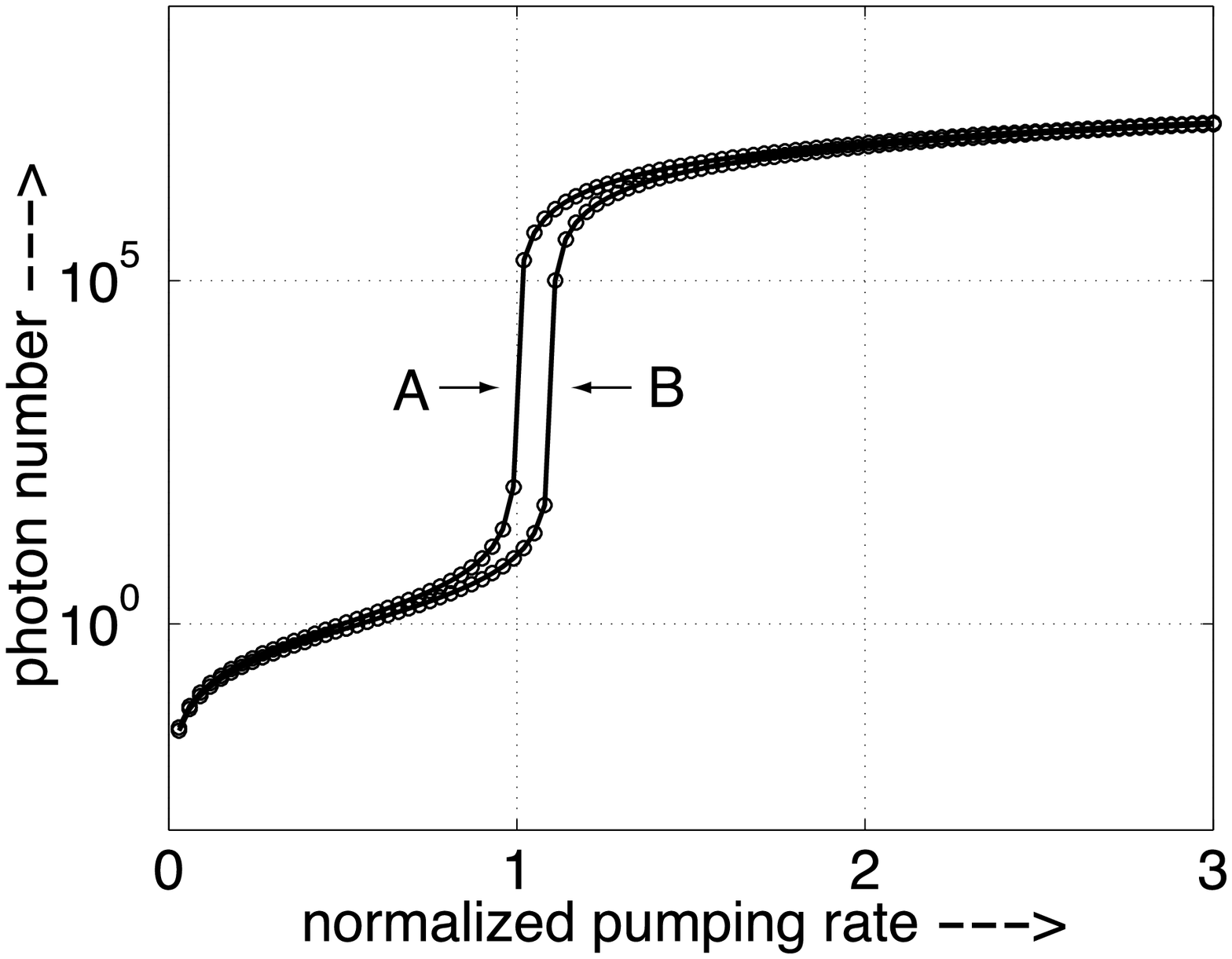}
\hspace{0.35in} 
\includegraphics[width=3in]{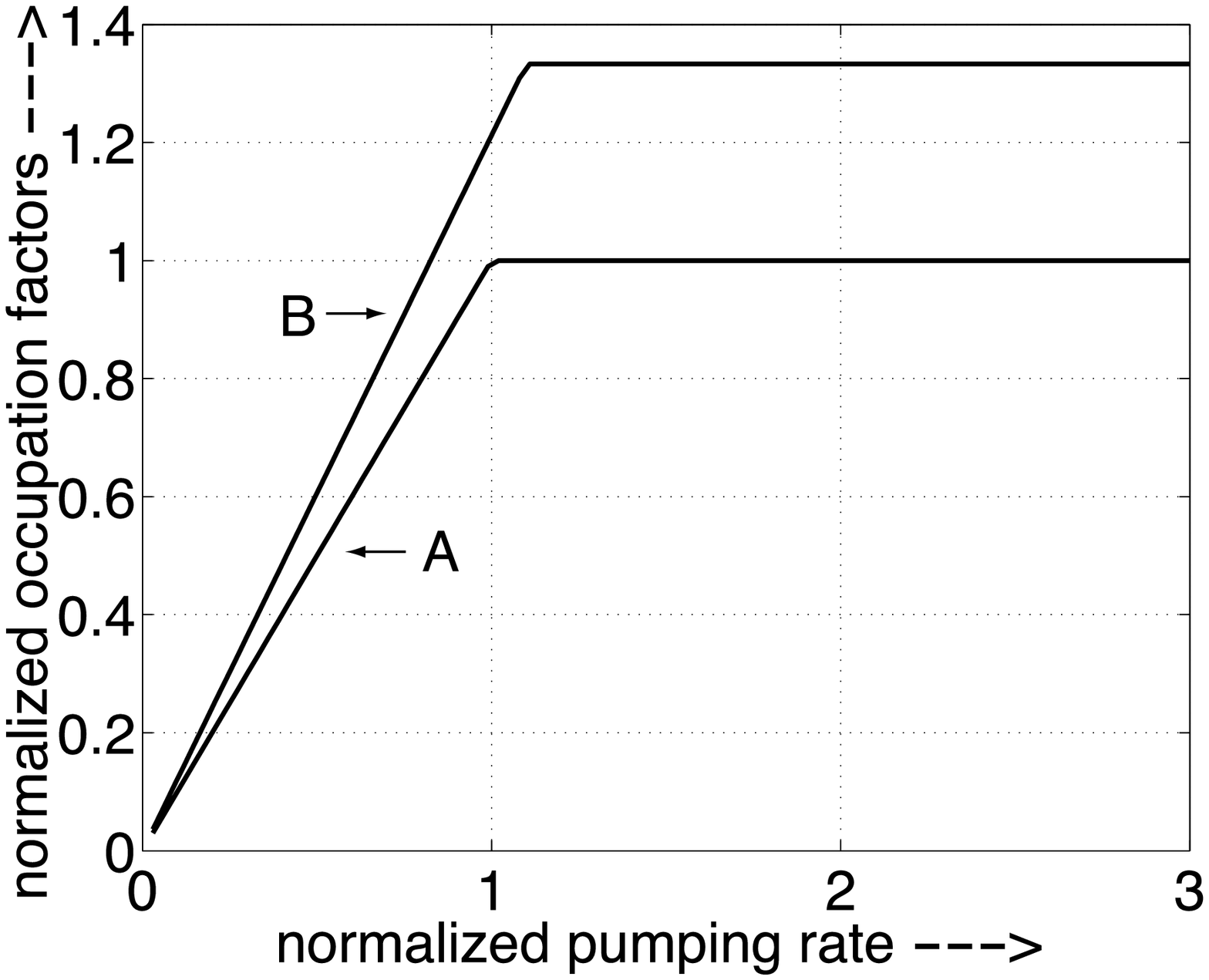}
\caption{(a) Photon numbers $n_A$ and $n_B$ and (b) normalized occupation
factors $f_A/f_{th}^A$ and $f_B/f_{th}^A$ when the electron scattering
rate between states $A$ and $B$ is $s=0$. 
The iterative
solution of Eq.~(\ref{sspectralmc}) (solid lines) matches the
analytical solutions for two independent single mode lasers 
(circles) having
$s=0$ from Eqs.~(\ref{nA0})-(\ref{nB0}).}
\label{mcs=0}
\end{figure}

\begin{figure}[htb]
\begin{center}
\includegraphics[width=3in]{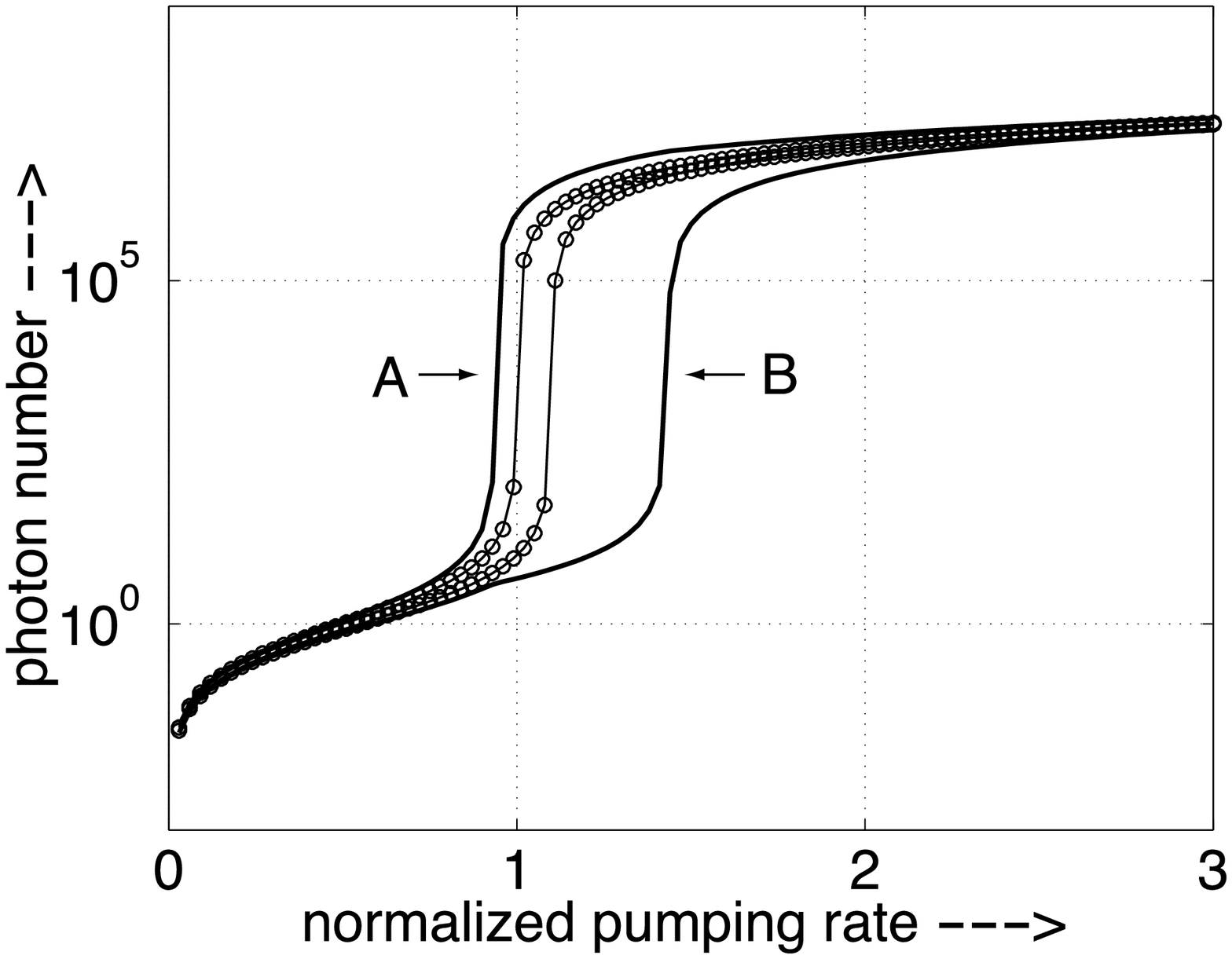}
\hspace{0.35in} 
\includegraphics[width=3in]{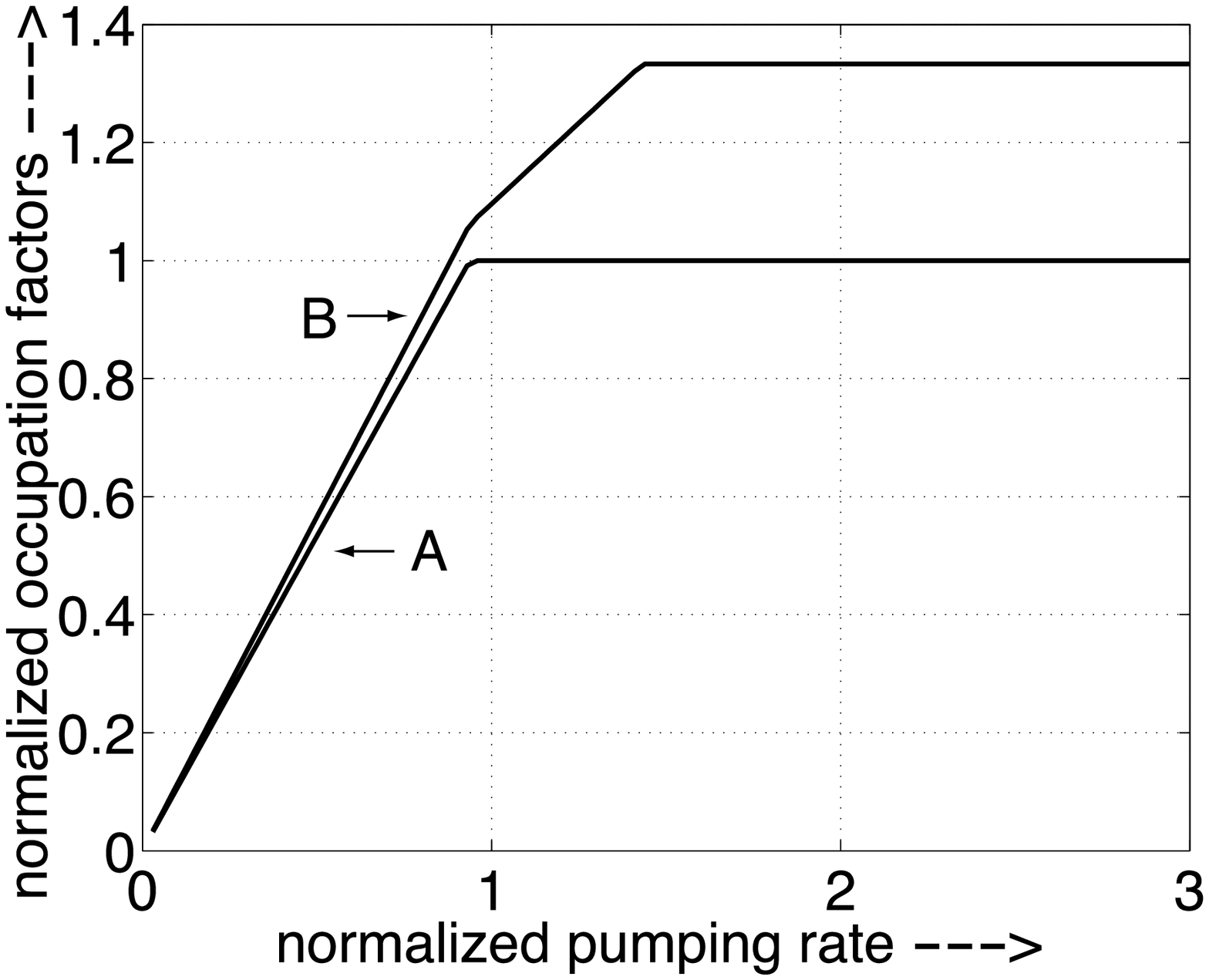}
\end{center}
\caption{(a) Photon numbers $n_A$ and $n_B$ and (b) normalized occupation
factors $f_A/f_{th}^A$ and $f_B/f_{th}^A$ when the electron scattering
rate $s$ between states $A$ and $B$ equals the spontaneous emission rate
of mode $A$ ($s=1.0A_A$). Scattering forces the occupation factors
$f_A$ and $f_B$ towards each other, reducing the threshold current for
mode $A$ and increasing the threshold current of mode $B$.}
\label{mcs=1.0AA}
\end{figure}

\begin{figure}[htb]
\includegraphics[width=3in]{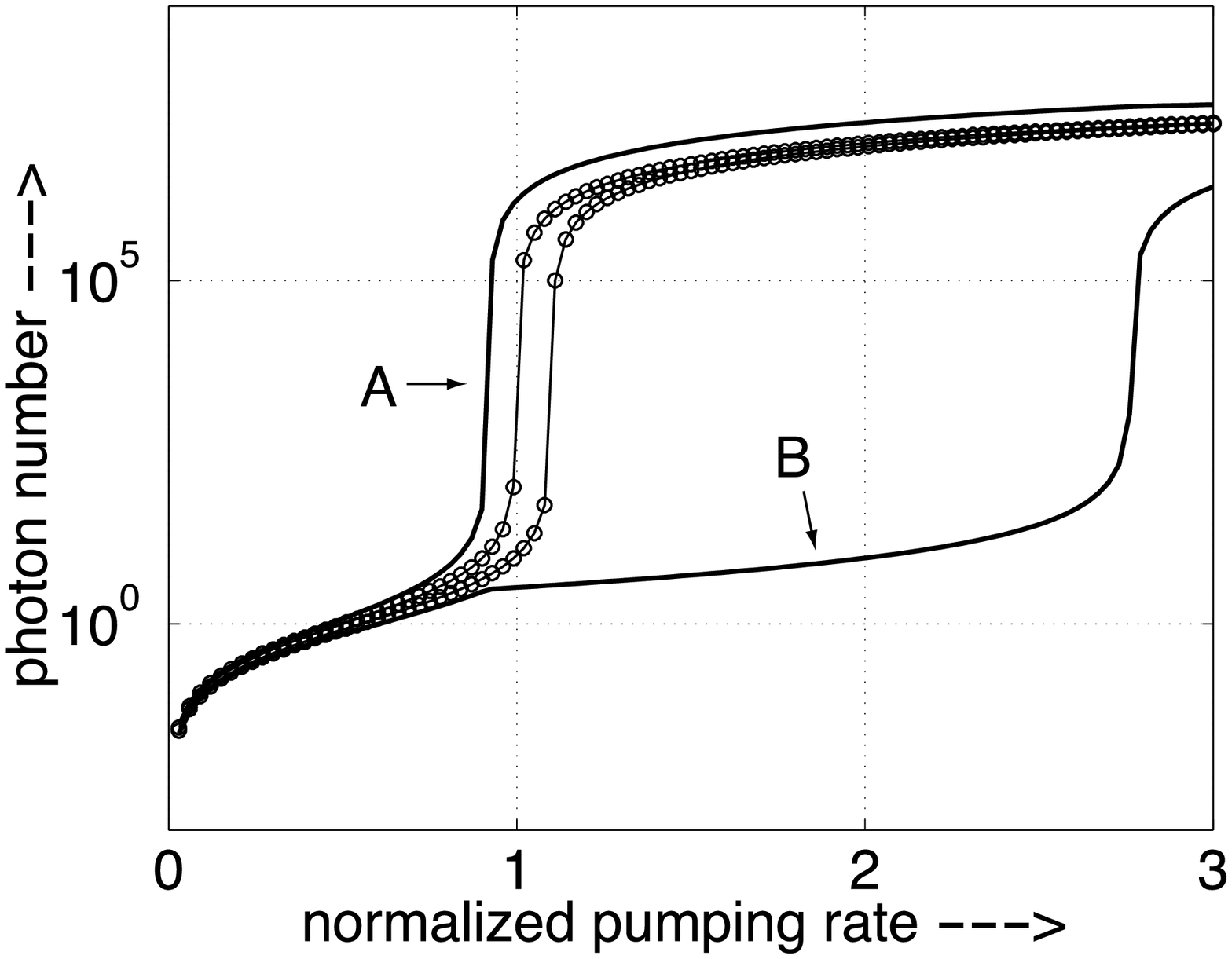}
\hspace{0.35in} 
\includegraphics[width=3in]{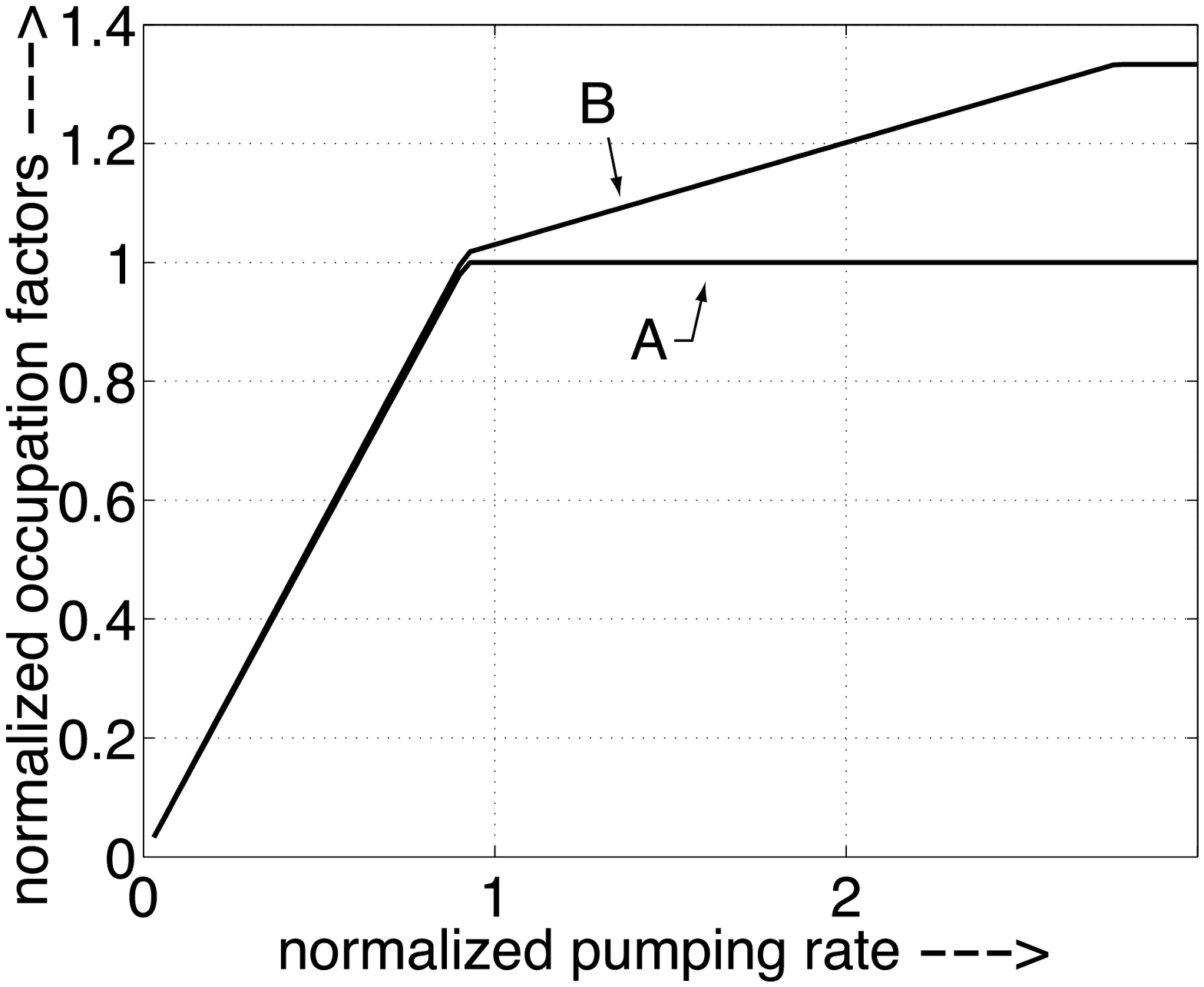}
\caption{(a) Photon numbers $n_A$ and $n_B$ and (b) normalized occupation
factors $f_A/f_{th}^A$ and $f_B/f_{th}^A$ when the electron scattering
rate $s$ between states $A$ and $B$ is five times the spontaneous emission rate
of mode $A$ ($s=5.0A_A$). The threshold current of mode $A$ decreases slightly
from Fig.~\ref{mcs=1.0AA}, 
while the threshold current for mode $B$ substantially increases.}
\label{mcs=5.0AA}
\end{figure}

We use some results from the single mode laser~\cite{siegman71,siegman86} 
to frame our discussion of two coupled lasing modes. When the two laser 
modes are decoupled ($s=0$), the threshold currents for each mode are
$I_{th}^A = \gamma_A A_A / N_A K_A$ and $I_{th}^B = \gamma_B A_B / N_B K_B$. 
The normalized pumping rate $r$ is defined as $r = I_A / I_{th}^ A$. 
We define the ratio of the two threshold currents when the modes are
decoupled as $z = I_{th}^B / I_{th}^A$. We assume the pumping current 
divides equally among the two states so that $I_A = I_B = I/2$. The occupation
factors above threshold are fixed at $f_A = f_{th}^A = \gamma_A/N_A K_A$
and $f_B = f_{th}^B = \gamma_B/N_B K_B$ due to gain saturation.

An important laser parameter 
is the number of luminescent modes $p$, where $A_A \equiv p K_A$. 
For mode $B$ this leaves the relation 
$A_B \equiv p z K_B (\gamma_A/\gamma_B)(N_B/N_A)$. We further
simplify by taking $N = N_A = N_B$ and $\gamma_A = \gamma_B = \gamma$. 
In the absence of mode coupling ($s=0$), the results for photon numbers versus
normalized pumping rate $r$ are then
\begin{equation}
2 n_{A} = p(r-1) + p \sqrt{(r-1)^2 + (4r/p)}
\label{nA0}
\end{equation}
and
\begin{equation}
2 n_{B} = p(r-z) + p \sqrt{(r-z)^2 + (4r/p)}.
\label{nB0}
\end{equation}

Figures~\ref{mcs=0}-\ref{mcs=5.0AA} show the photon numbers and occupation
factors versus normalized pumping rate $r$ for different scattering rates $s$.
We choose $p = 10^7$ and $z=1.1$ in Figs.~\ref{mcs=0}-\ref{mcs=5.0AA}.
In Fig.~\ref{mcs=0} there is no scattering between states $A$ and $B$ 
($s=0$), leaving two independent single mode lasers. Iterating
Eq.~(\ref{iterate}) (solid lines) then just reproduce the analytical results
from Eqs.~(\ref{nA0})-(\ref{nB0}) (circles) in Fig.~\ref{mcs=0}(a). The 
normalized occupation factors $f_A/f_{th}^A$ and $f_B/f_{th}^A$ 
in Fig.~\ref{mcs=0} increase
approximately linearly with pumping below threshold and saturate above the
lasing threshold.

As the scattering rate
increases to $s=1.0A_A$ in Fig.~\ref{mcs=1.0AA}, and even further 
to $s=5.0A_A$ in Fig.~\ref{mcs=5.0AA},
we see the lasing threshold for mode $A$ shifts to a lower pumping current.
The threshold current for mode $B$ continues to increase as the scattering rate
$s$ increases. The increase in lasing threshold for mode $B$ is much more
pronounced than the decrease in threshold current for mode $A$. Inspection of the
occupation factors for the two decoupled lasers in Fig.~\ref{mcs=0}(b) explains
the threshold current shifts. Increasing the scattering rate $s$ forces 
the two occupation factors towards each other. 
In Fig.~\ref{mcs=0}(b) we have $f_B>f_A$ in the subthreshold region.
Hence scattering between the modes will increase $f_A$ and
decrease $f_B$ in the subthreshold region as seen in Fig.~\ref{mcs=1.0AA}(b)
and Fig.~\ref{mcs=5.0AA}(b). Scattering then lowers the threshold current required for
mode $A$ to lase. Once mode $A$ reaches the lasing threshold, additional
scattering between states $A$ and $B$ makes it more difficult for mode $B$ to raise
its occupation factor to $f_B = f_{th}^B$ required for mode $B$ to lase.

The threshold current for mode $A$ can shift in either direction, up or down, 
with additional scattering between the modes. 
If $f_B > f_A$ in the subthreshold region, the case we have chosen
in Figs~\ref{mcs=0}-\ref{mcs=5.0AA}, additional scattering $s$ lowers
the threshold current for mode $A$. If the occupation factors obey $f_B < f_A$ 
in the subthreshold region, then the threshold current for mode $A$ increases
with additional scattering $s$. When the pumping current divides equally between
the states $A$ and $B$ as we have assumed, the threshold current 
for mode $A$ shifts down with
additional scattering $s$ if $(f_{th}^B/I_{th}^B) > (f_{th}^A/I_{th}^A)$, or,
equivalently,  
if the spontaneous rates obey $A_A > A_B$. Given
our assumptions of $\gamma_A= \gamma_B = \gamma$ and $N_A = N_B = N$, we
require $K_A > z K_B$ for additional scattering $s$ to lower the threshold
current of mode $A$. Since we choose $z=1.1$ and $K_B = 0.75 K_A$ in 
Figs.~\ref{mcs=0}-\ref{mcs=5.0AA} 
this condition is satisfied. Increasing $z$ and/or $K_B$ could reverse the
inequality and raise the threshold current for mode $A$ with increased scattering $s$.
If the
pumping current divides unequally as $I_A = \alpha I$ and $I_B = (1-\alpha) I$, the
requirement for additional scattering $s$ to lower the threshold current of mode
$A$ is $(1-\alpha) A_A > \alpha A_B $. 
The threshold current required for mode $B$ to lase will always increase 
when we add additional scattering $s$ (assuming mode $A$ is the favored lasing
mode with $f_{th}^A < f_{th}^B$).

\begin{figure}[htb]
\includegraphics[width=3in]{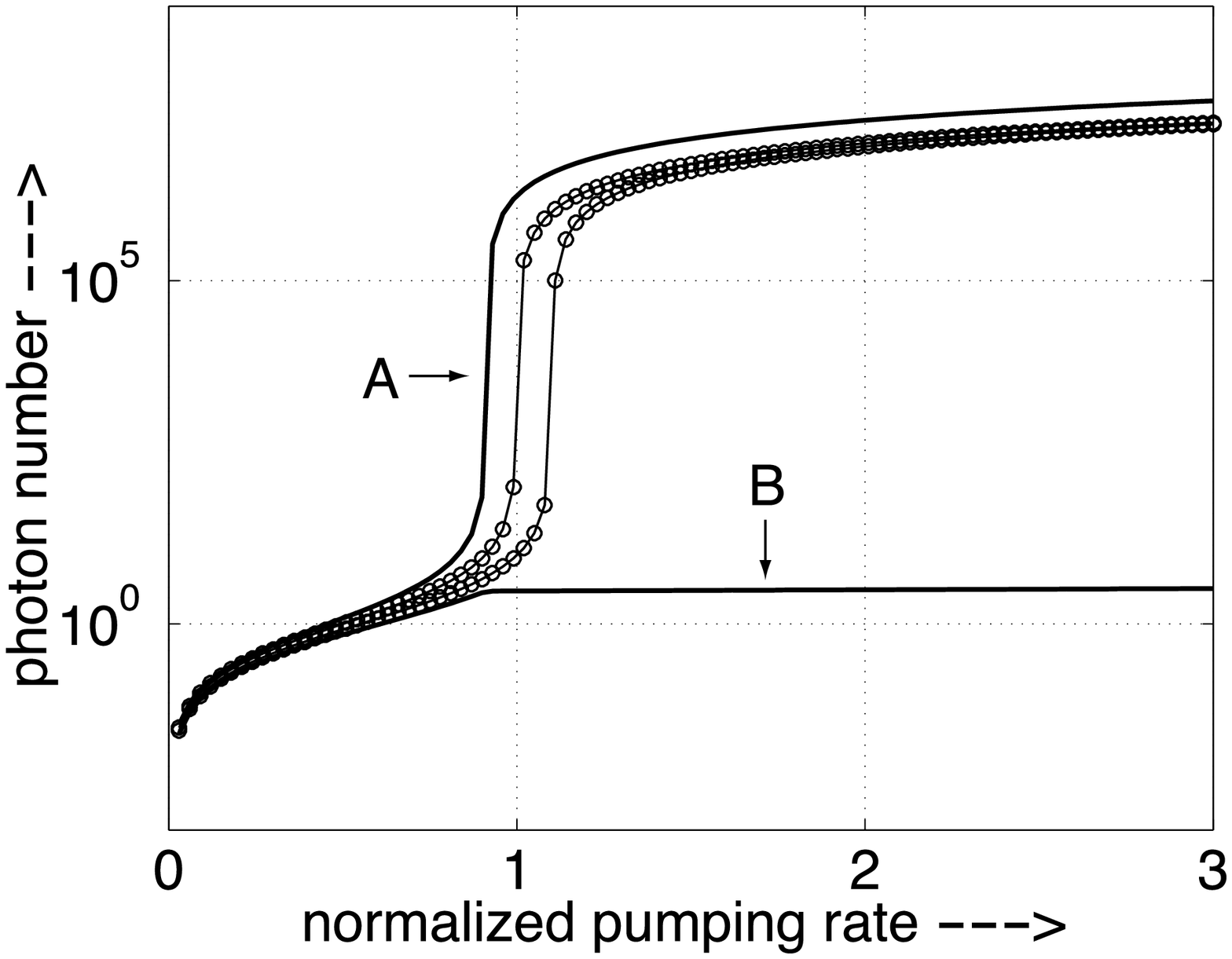}
\hspace{0.35in} 
\includegraphics[width=3in]{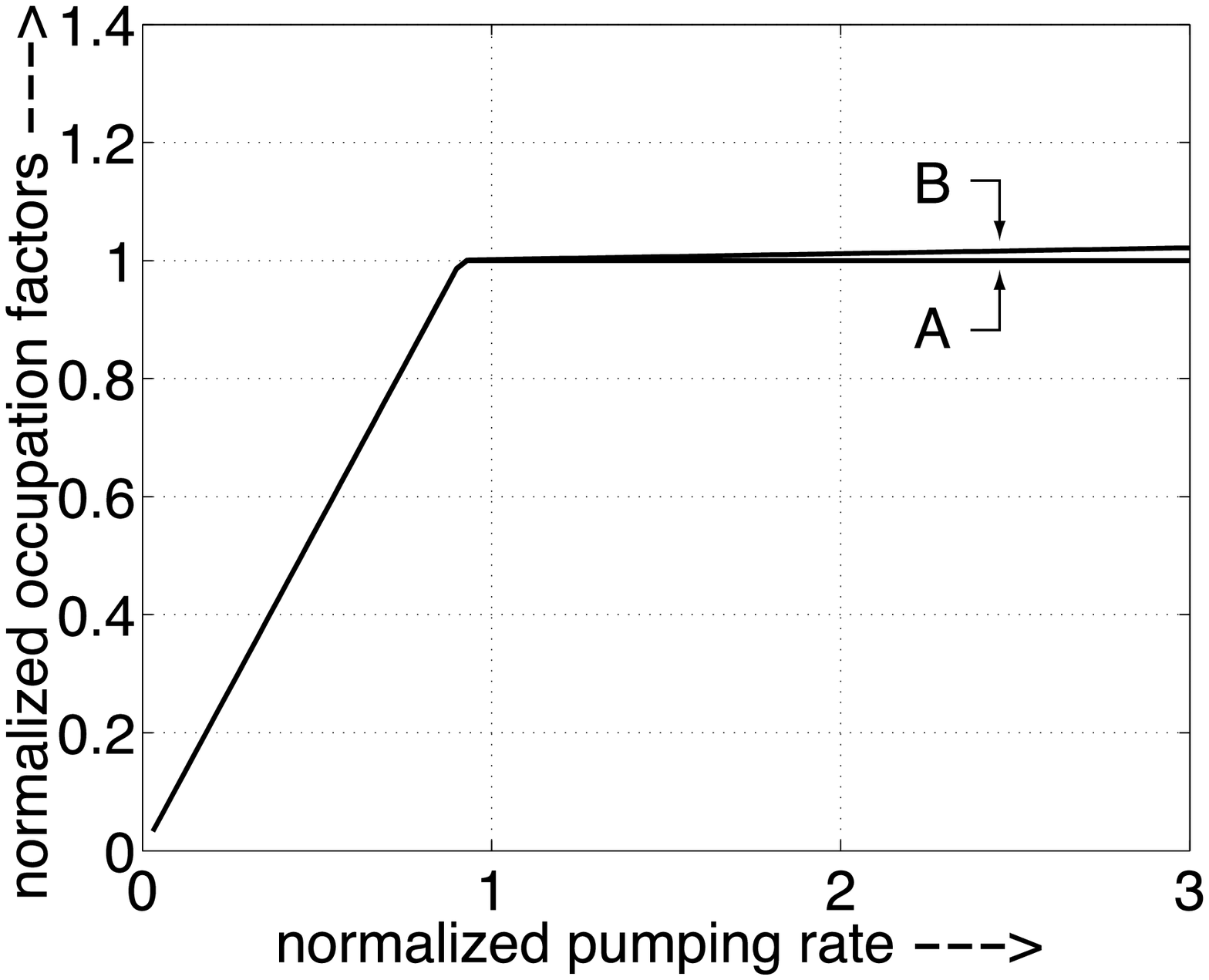}
\caption{(a) Photon numbers $n_A$ and $n_B$ and (b) normalized occupation
factors $f_A/f_{th}^A$ and $f_B/f_{th}^A$ when the electron scattering
rate $s$ between states $A$ and $B$ approaches infinity ($s=100A_A$). 
Since $s \to \infty$ we approach the limit of a homogeneous laser line
and of a single mode laser.}
\label{homog}
\end{figure}

In the limit of $s \to \infty$ we move towards a homogeneous line. 
Figure~\ref{homog} shows the solution of Eq.~(\ref{sspectralmc}) with
with $s = 100 A_A$. The photon number $n_B$ and occupation factor $f_B$
are now essentially fixed when mode $A$ starts lasing due to gain saturation. 
Iteratively solving Eq.~(\ref{homogline}) 
produces essentially the same graph as shown in Fig.~\ref{homog}. The
homogeneous line shown in Fig.~\ref{homog} is the opposite limit of two
independent laser lines shown in Fig.~\ref{mcs=0}. Varying the scattering
rate $s$ interpolates smoothly between the solutions in 
Fig.~\ref{mcs=0} and Fig.~\ref{homog}.

\clearpage

\section{Semiconductor Lasers: Spatial Hole Burning}
\indent

\label{semilaser}

\begin{figure}[htb]
\includegraphics[width=6in]{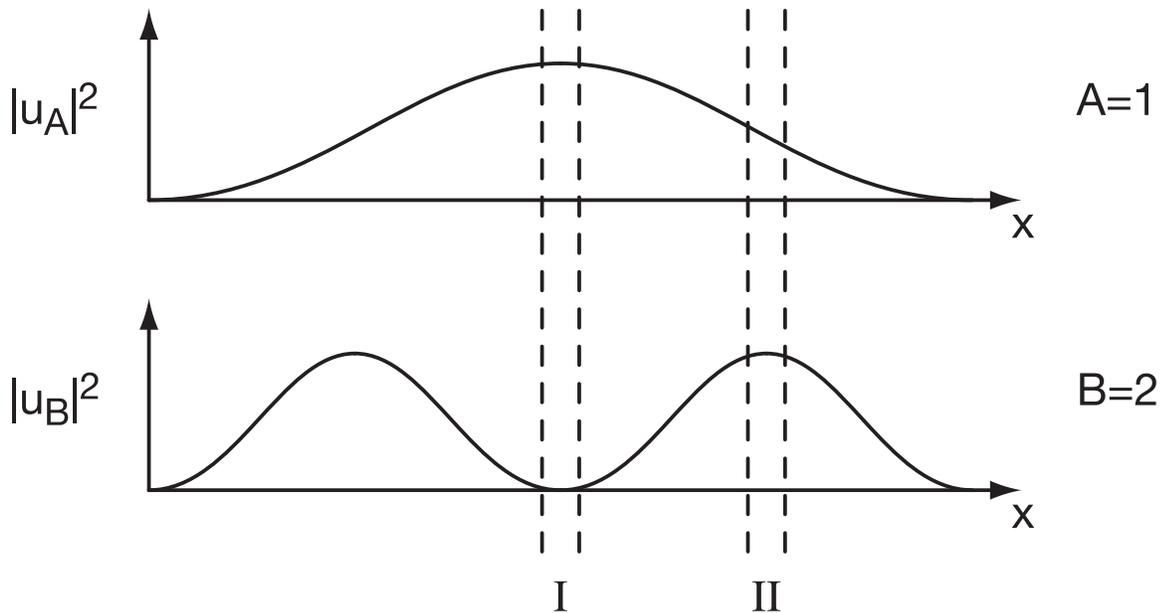}
\caption{Competition between two lasing modes $A$ and $B$ is possible
on a homogeneous line when the optical mode intensities $|u_A|^2$
and $|u_B|^2$ vary in space. Mode $A$ is favored in region I, while
in region II mode $B$ is the favored lasing mode.}
\label{qualsem}
\end{figure}

When there are spatial variations in the optical intensity in different modes,
two lasing frequencies can coexist on a homogeneous line.
Figure~\ref{qualsem} shows the normalized optical mode intensities $|u_A|^2$
and $|u_B|^2$ for the lowest two longitudinal modes in a cavity.
Near the center of the laser (region I), mode $A$ is the favored lasing mode.
However in region II, where the optical intensity $|u_B|^2 > |u_A|^2$,
mode $B$ is the favored lasing mode. If the gain medium were confined to
region I, only mode $A$ would lase. Similarly, for the gain medium restricted
to region II, only mode $B$ would lase. For semiconductor lasers the
gain media fills the entire laser cavity, so there is competition 
for the available optical gain between the lasing modes.

Whether or not a single or multiple frequencies appear in the laser output
spectrum depends on the size of the electron diffusion coefficient $D$. For 
single frequency laser operation to occur the electron must diffuse from 
region I to region II  in Fig.~\ref{qualsem} before the photon exits the 
cavity. If the photon escapes the laser cavity before the electron can
diffuse from region I to region II, the regions are essentially independent
as far as the laser light is concerned. Optically, the laser behaves as if two
independent (single mode) lasers operate inside the cavity. The distance from 
region I to region II  in Fig.~\ref{qualsem} is approximately one quarter of 
the lasing wavelength ($\lambda/4$). So for open cavity lasers we expect essentially
single mode operation whenever $D \gg (\lambda/4)^2 \tilde{A}$.
If the cavity is closed (no side luminescence) so that $\tilde{A} \to \gamma_A$, 
the condition for single mode laser operation becomes $D \gg (\lambda/4)^2 \gamma_A$. 

To describe semiconductor lasers quantitatively we 
need to generalize Eq.~(\ref{spectralmc}) to account for the
spatial variation in the mode patterns and electron density 
inside the laser. The occupation factors in each mode will be
spatially varying such that $f_A \to f_A(r)$ and 
$f_B \to f_B(r)$. If we introduce the position dependent
density of states $N(E,r)$, the total electron density is now
\begin{equation}
\rho(r,t) = \sum_i N(E_i,r) f_i(r,t).
\label{pdos}
\end{equation}
The electron density in states $A$ and $B$ are 
$\rho_A(r) = N(E_A,r) f_A(r)$ and $\rho_B(r) = N(E_B,r) f_B(r)$.
The total number of active type $A$ lasing levels is then
\begin{equation}
N_{A} = \int_a N(E_A,r) dV,
\end{equation}
where $\int_a$ denotes integration over that portion of the laser
cavity containing the active lasing media.
We further define the scattering rate per initial and final state density 
$\tilde{s}_{AB}$ as
\begin{equation}
\tilde{s}_{AB} = \frac{s_{AB}}{N(E_A,r) N(E_B,r)}.
\end{equation}

To account for spatial variations in the electromagnetic modes
inside the laser cavity we introduce the mode functions $u_A(r)$ and
$u_B(r)$ such that the electromagnetic energy density is given
\begin{equation}
\hbar \omega_A n_A(t) |u_A(r)|^2 = \epsilon(r) |E_A(r,t)|^2,
\end{equation}
where $E_A(r,t)$ is the electric field of mode $A$ and $\epsilon(r)$
the dielectric constant. Since we must have 
\begin{equation}
\int_L \epsilon(r) |E_A(r,t)|^2 dV = \hbar \omega_A n_A(t) ,
\end{equation}
where the integration region $L$ denotes the entire laser cavity,
the mode functions $u_A(r)$ are normalized as
\begin{equation}
\int_L |u_A(r)|^2 dV = 1.
\label{normalize}
\end{equation}
We can insert this factor of '1' from Eq.~(\ref{normalize}) wherever necessary
in order to generalize Eq.~(\ref{spectralmc}) to account for spatially varying
electromagnetic fields.

Using Eqs.~(\ref{pdos})-(\ref{normalize}), the generalization of Eq.~(\ref{fa}) 
to account for spatial variations in
the electron density and electromagnetic field intensity is
\begin{eqnarray}
\frac{d \rho_A}{dt} = N(E_A,r) 
\left\{ - \rho_A \tilde{s}_{AB} [N(E_B,r) -\rho_B] 
+ \rho_B \tilde{s}_{BA} [N(E_A,r) -\rho_A] \right\} \nonumber \\
+ R_A(r,t) - [K_A V_L] n_A |u_A|^2 \rho_A
- A_A \rho_A + D_A \nabla^2 \rho_A .
\label{rhoa}
\end{eqnarray}
Here $R_A(r,t) = N(E_A,r) I_A(r,t)$ is the total pumping rate per unit volume
into the state $A$, $V_L = \int_L dV$ the volume of the laser cavity, and
$D_A$ the diffusion constant of electrons in state $A$.
The generalization of Eq.~(\ref{na}) to account for spatial variations 
inside the laser is
\begin{equation}
\frac{dn_A}{dt} = [K_A V_L] (n_A+1) \int_a |u_A|^2 \rho_A dV - \gamma_A  n_A.
\label{nra}
\end{equation}
Eqs.~(\ref{rhoa})-(\ref{nra}) can be used to construct a generalization 
of the coupled mode Eq.~(\ref{spectralmc}) to account for spatial variations in 
the laser.

Our interest is in semiconductors with homogeneous optical lines, so we do not
pursue the full generalization of Eq.~(\ref{spectralmc}). We assume the
scattering rate $s_{AB} \to \infty$ in the semiconductor, so that we are
back on a homogeneous optical line. We assume negligible separation of the
energy levels as before so that $(E_A-E_B) \ll k_B T$ and $s = s_{AB} = s_{BA}$
The occupation factors we therefore take to be in equilibrium with each other 
at each point in space so that $f(r) = f_A(r) = f_B(r)$. With these assumptions 
we have
\begin{eqnarray}
\frac{d \rho(r,t)}{dt} = 
- \left( [K_A V_L] n_A(t) |u_A(r)|^2 + A_A \right)
\frac{N(E_A,r)}{N(E_A,r)+N(E_B,r)}\rho(r,t)
\nonumber \\
- \left( [K_B V_L] n_B(t) |u_B(r)|^2 + A_B \right)
\frac{N(E_B,r)}{N(E_A,r)+N(E_B,r)}\rho(r,t)
\nonumber \\
+ D \nabla^2 \rho(r,t) + R(r,t).
\end{eqnarray}
Here $\rho = \rho_A + \rho_B$ is the total electron density,
the total pumping rate is $R=R_A+R_B$, and we have taken $D=D_A=D_B$ for
the diffusion constant. The final coupled mode rate equations for a 
homogeneous semiconductor line that we solve are 
\begin{equation}
\frac{d \rho}{dt} = 
- [\tilde{K}_A V_L] n_A |u_A|^2 \rho
- [\tilde{K}_B V_L] n_B |u_B|^2 \rho
-\tilde{A} \rho + D \nabla^2 \rho + R,
\label{homorho}
\end{equation}
together with 
\begin{equation}
\frac{dn_A}{dt} = (n_A+1) \int_a [\tilde{K}_A V_L] |u_A|^2 \rho dV - \gamma_A  n_A,
\label{nrat}
\end{equation}
and
\begin{equation}
\frac{dn_B}{dt} = (n_B+1) \int_a [\tilde{K}_B V_L] |u_B|^2 \rho dV - \gamma_B  n_B.
\label{nrbt}
\end{equation}
The position dependent optical rate constants $\tilde{K}_A(r)$, 
$\tilde{K}_B(r)$, and $\tilde{A}(r)$ are
\begin{equation}
\tilde{K}_A(r) = K_A \frac{N(E_A,r)}{N(E_A,r)+N(E_B,r)},
\label{tka}
\end{equation}
with
\begin{equation}
\tilde{K}_B(r) = K_B \frac{N(E_B,r)}{N(E_A,r)+N(E_B,r)},
\label{tkb}
\end{equation}
and
\begin{equation}
\tilde{A}(r) = 
\frac{A_A N(E_A,r) + A_B N(E_B,r)}{N(E_A,r)+N(E_B,r)}.
\label{ta}
\end{equation}

Eqs.~(\ref{homorho})-(\ref{nrbt}) are similar to Eqs.~(E.1.9a) and (E.1.9b)
for a single mode laser from Ref.~\cite{svelto}. Eqs.~(\ref{homorho})-(\ref{nrbt}) 
should also be considered the generalization of  Eq.~(\ref{homogline}) 
to account for spatial variations while lasing on a homogeneous line.
We simplify further by taking the electron density
of states to be constant in space so that the rate constants $\tilde{K}_A(r)$,
$\tilde{K}_B(r)$, and $\tilde{A}(r)$ are independent of space. 
For simplicity and concreteness 
we consider competition between the longitudinal
laser modes, though the same procedure would work for the inclusion of
transverse cavity modes. We choose an Fabry-Perot type cavity having
the normalized mode functions
\begin{equation}
|u_A|^2 = \frac{2}{A_L L} \sin^2(A \pi x/L)
\end{equation}
and 
\begin{equation}
|u_B|^2 = \frac{2}{A_L L} \sin^2(B \pi x/L).
\end{equation}
Here $A_L$ is the cavity area, $L$ the cavity length,
$A$ is the number of half wavelengths in mode $A$, and $B$ the number of
half wavelengths in the longitudinal cavity mode $B$. 

\subsection{Slow Diffusion}
\indent

We solve Eqs.~(\ref{homorho})-(\ref{nrbt}) in steady state using an iterative
technique. We take the diffusion constant $D=0$, letting
us solve Eq.~(\ref{homorho}) for $\rho$ and substitute
back into Eqs.~(\ref{nrat})-(\ref{nrbt}) leading to
\begin{equation}
n_A = (n_A + 1) p \int_a 
\frac{|u_A|^2 r (dV/V_a)}
{ |u_A|^2 n_A + |u_B|^2 n_B (1/z) + (p/V_L)},
\label{nrat2}
\end{equation}
and
\begin{equation}
n_B = (n_B + 1) p \int_a 
\frac{(1/z) |u_B|^2 r (dV/V_a)}
{ |u_A|^2 n_A + |u_B|^2 n_B (1/z)  + (p/V_L)}.
\label{nrbt2}
\end{equation}
In Eqs.~(\ref{nrat2})-(\ref{nrbt2}) we have used the ration of optical coupling
constants $z=\tilde{K}_A/\tilde{K}_B$, the number of luminescent modes
$p=\tilde{A}/\tilde{K}_A$, assumed equal cavity escape rates $\gamma_A = \gamma_B$,
and defined the normalized pumping rate as $r = R/R_{th}^A = R V_a / \gamma p$.
We now produce the
$(m+1)$st iteration for the photon numbers from the $m$th iteration using
\begin{equation}
n_A(m+1) = p \int_{L_a}
\frac{(n_A(m) + 1) r(x) \sin^2(A \pi x/L) (dx/L_a)}
{ n_A(m) \sin^2(A \pi x/L) + (1/z) n_B(m) \sin^2(B \pi x/L) + (p/2)},
\label{nram}
\end{equation}
and
\begin{equation}
n_B(m+1) = p \int_{L_a}
\frac{(n_B(m) + 1) r(x) (1/z) \sin^2(B \pi x/L) (dx/L_a)}
{ n_A(m) \sin^2(A \pi x/L) + (1/z) n_B(m) \sin^2(B \pi x/L) + (p/2)}.
\label{nrbm}
\end{equation}
Here $V_a = A_L L_a$ with $A_L$ the cavity area and $L_a$ the length of
active media. Once we have iterated Eqs.~(\ref{nram})-(\ref{nrbm}) to
convergence, we obtain the electron density from
\begin{equation}
\frac{\rho(x)}{\rho_{th}^A} = \frac{p r(x)}
{p + 2 n_A \sin^2(A \pi x/L) + (2/z) n_B \sin^2(B \pi x/L) }.
\end{equation}
Here $\rho_{th}^A = N_A f_{th}^A/V_a = \gamma/\tilde{K}_A V_a$ is the
electron density when mode $A$ reaches threshold in a single mode laser.
For simplicity we also take the pumping rate $r(x)$ to be a constant
(independent of space).

\begin{figure}[htb]
\includegraphics[width=3in]{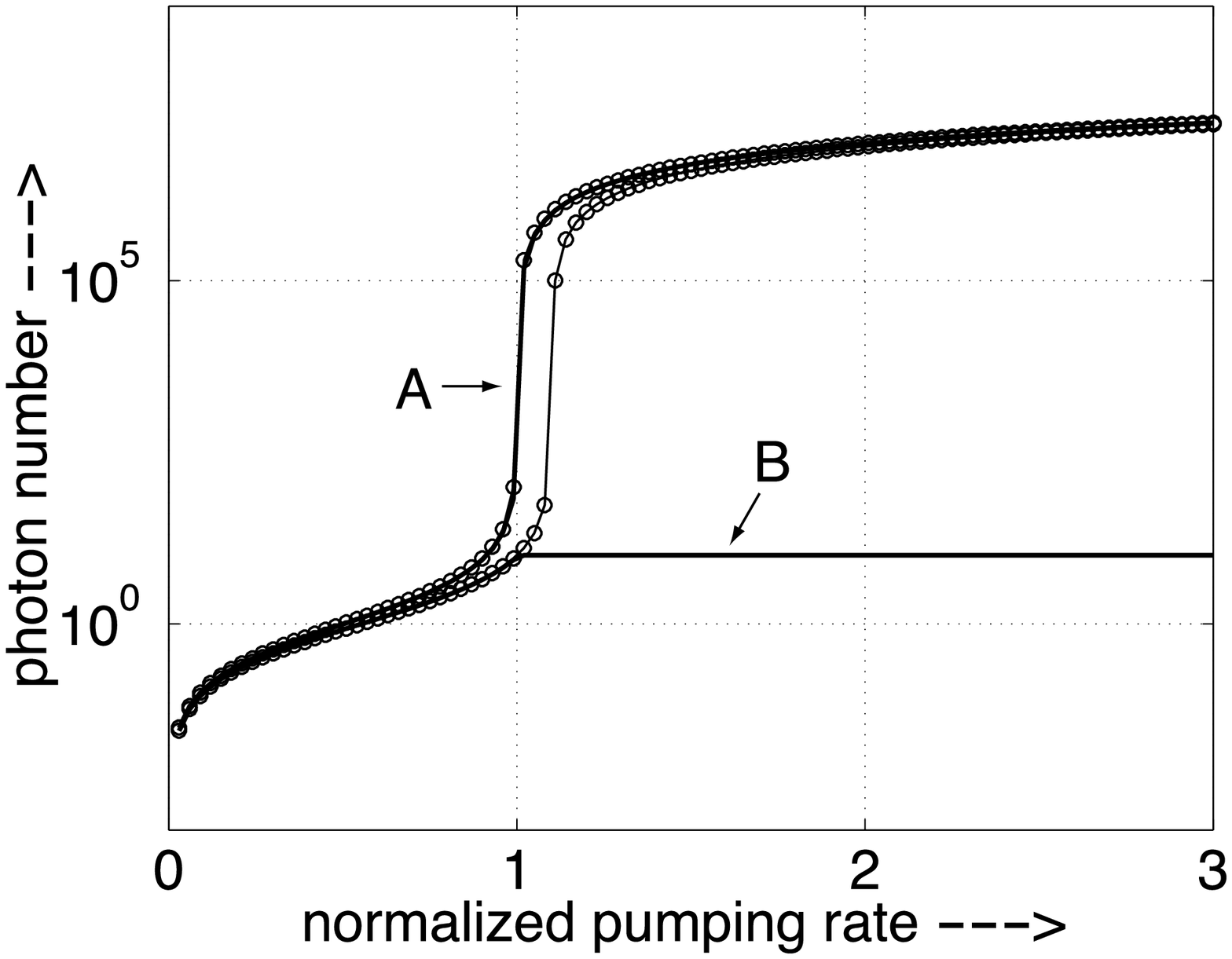}
\hspace{0.35in} 
\includegraphics[width=3in]{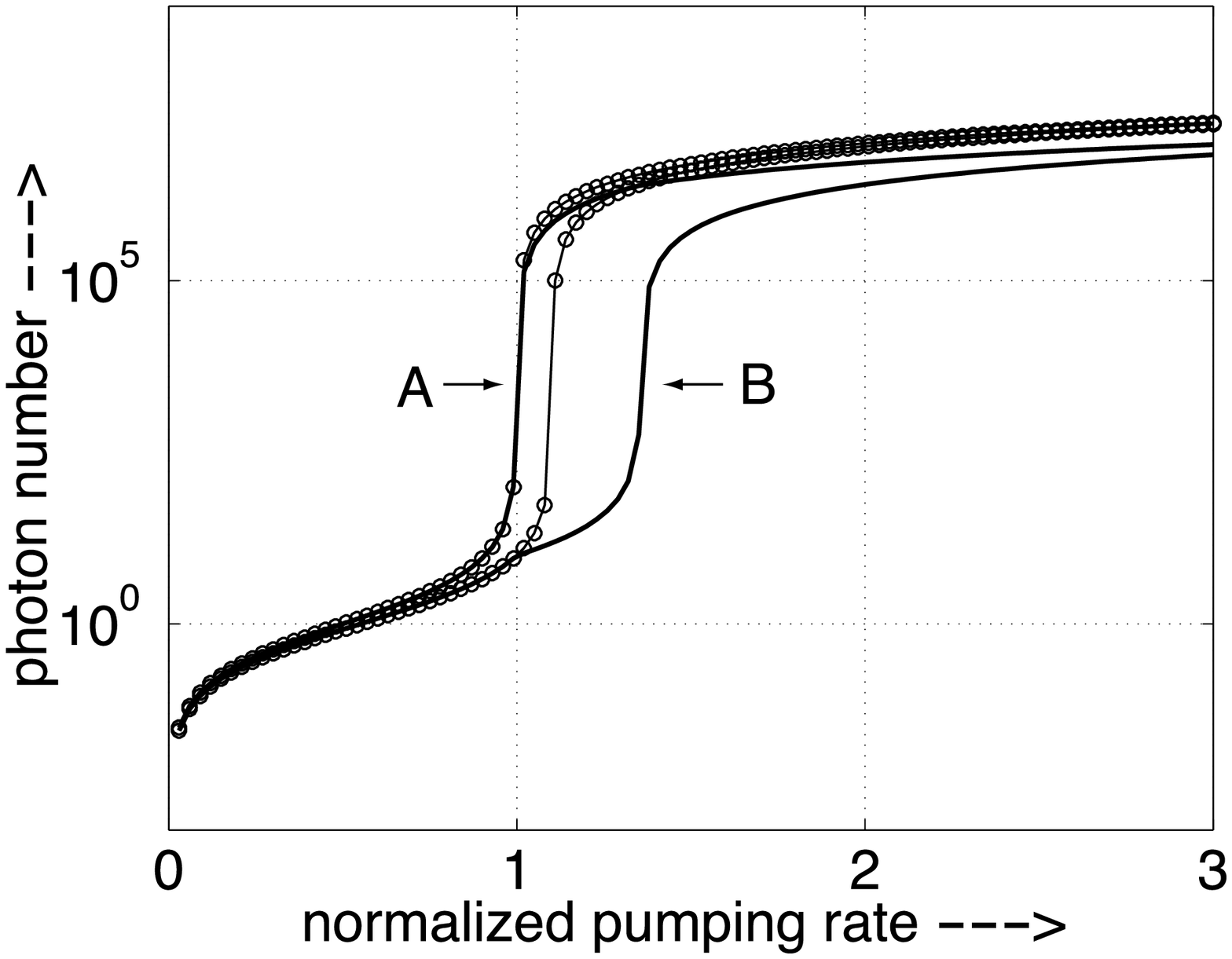}
\caption{Photon numbers $n_A$ and $n_B$ versus normalized pumping
rate $r$ when the optical mode intensities 
$|u_A|^2$ and $|u_B|^2$ are (a) constant in space
and (b) have spatial variation. Mode $B$ cannot lase when
the mode intensities are uniform in (a). Spatial variations 
in the mode intensities allow mode $B$ to lase in (b).}
\label{sema1b2}
\end{figure}

Figure~\ref{sema1b2} shows the photon numbers $n_A$ and $n_B$ for two
modes competing on a homogeneous line. In Fig.~\ref{sema1b2}(a) the optical
mode intensities are constant, so that $|u_A|^2 = |u_B|^2 = 1/ (A_L L)$ (as
we implicitly assumed for the gas laser of section \ref{gaslaser}).
Figure~\ref{sema1b2}(a) therefore mimics the case where spatial variations
in the laser are negligible. Two other cases where we can neglect spatial variation
of the mode intensities are in a ring laser or in a semiconductor laser
with rapid electron diffusion. In Fig.~\ref{sema1b2}(a) the photon
number $n_B$ in mode $B$ is fixed whenever mode $A$ begins lasing. The solution
of Eqs.~(\ref{nrat2})-(\ref{nrbt2}) therefore reproduces lasing on a
homogeneous line whenever the optical mode intensities $|u_A|^2$ and
$|u_B|^2$ are constant. We let the optical mode intensities vary in
space in Fig.~\ref{sema1b2}(b), where we have taken the lowest two longitudinal
cavity modes ($A=1$ and $B=2$). Mode $B$ can indeed begin lasing 
in Fig.~\ref{sema1b2}(b), but requires a higher pumping rate than for 
two independent lasers on the same optical line.
We have chosen parameters $z=1.1$ and $p = 10^7$ in Fig.~\ref{sema1b2}. 
The circles in Fig.~\ref{sema1b2} show the solutions from 
Eqs.~(\ref{nA0})-(\ref{nB0}) for two independent single mode lasers.

\begin{figure}[htb]
\includegraphics[width=3in]{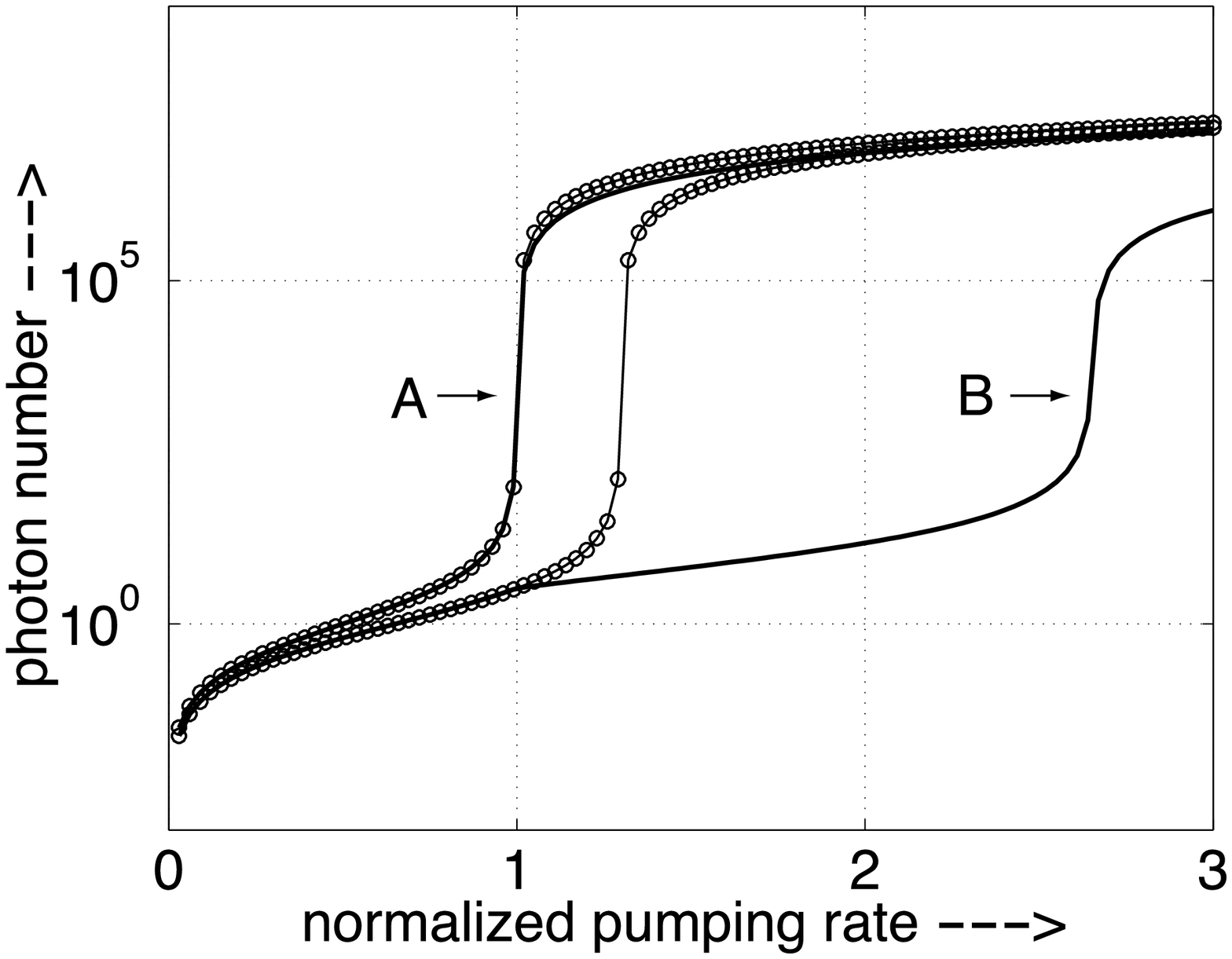}
\hspace{0.35in} 
\includegraphics[width=3in]{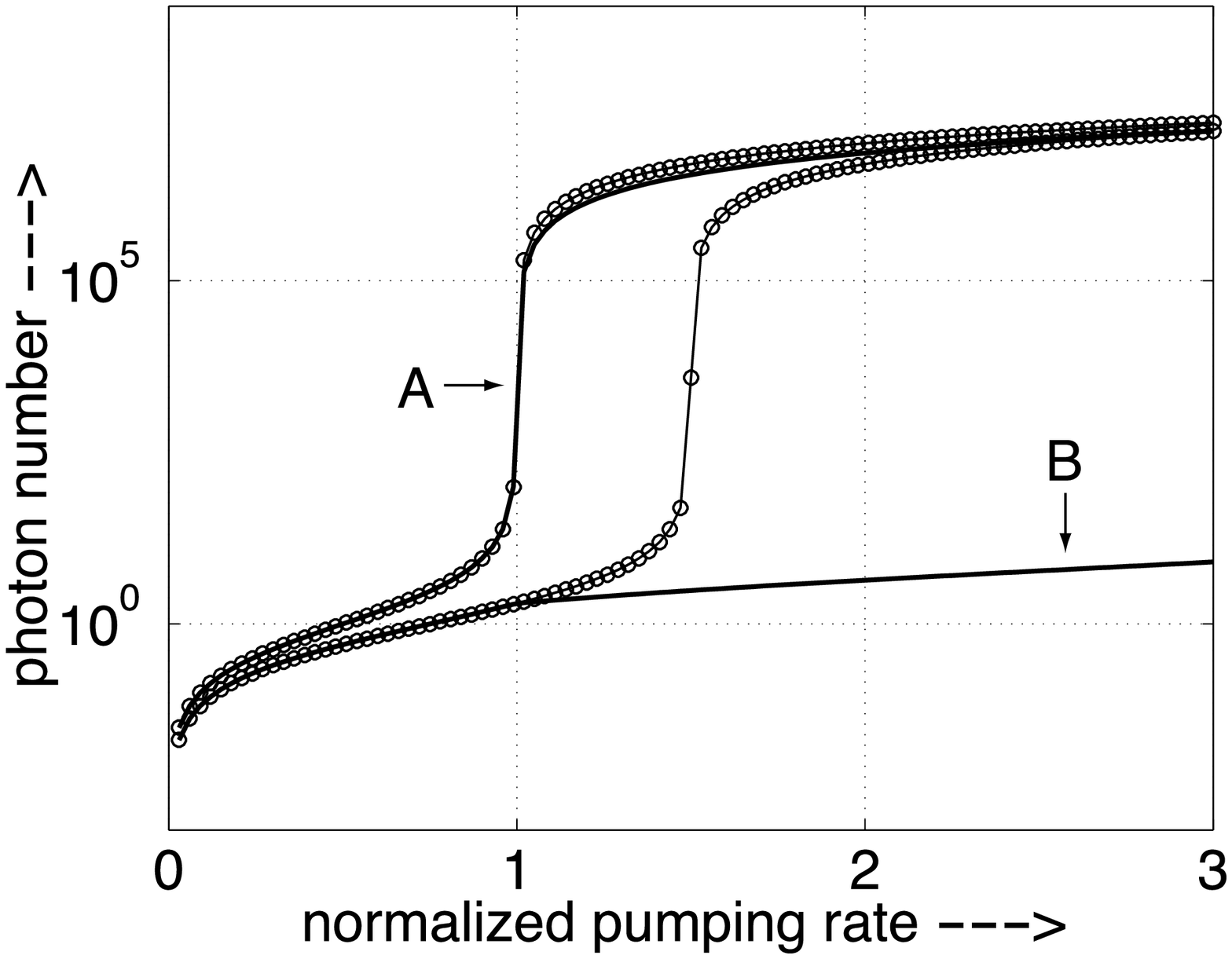}
\caption{Increasing the ratio of the optical rate constants 
from $z= \tilde{K}_A/\tilde{K}_B =1.1$ in Figure~\ref{sema1b2} 
to (a) $z=1.3$ and (b) $z=1.5$
increases the threshold current required for mode $B$ to lase. 
Spatial variations in the optical intensities
become less relevant when the optical rate constant for mode $B$
becomes too small.}
\label{semincz}
\end{figure}

Spatial vatiations in the optical mode intensities $|u_A|^2$ and 
$|u_B|^2$ become less relevant when the optical rate constant
$\tilde{K}_B$ becomes small. The ratio of the rate constants
$z = \tilde{K}_A/\tilde{K}_B$ in Fig.~\ref{sema1b2} is $z=1.1$.
We increase $z$ in Fig.~\ref{semincz} to (a) $z=1.3$ and (b) $z=1.5$,
raising the threshold current required for mode $B$ to lase. For the
parameter $z=1.5$ in Fig.~\ref{semincz}(b), mode $B$ no longer
lases for the range of pumping currents shown $(0 \le r \le 3)$.
Although spatial variations in the optical mode intensities are
still present in Fig.~\ref{semincz}, they become less relevant
when the optical coupling constant $\tilde{K}_B$ for mode $B$ 
is too weak.

\begin{figure}[htb]
\includegraphics[width=3in]{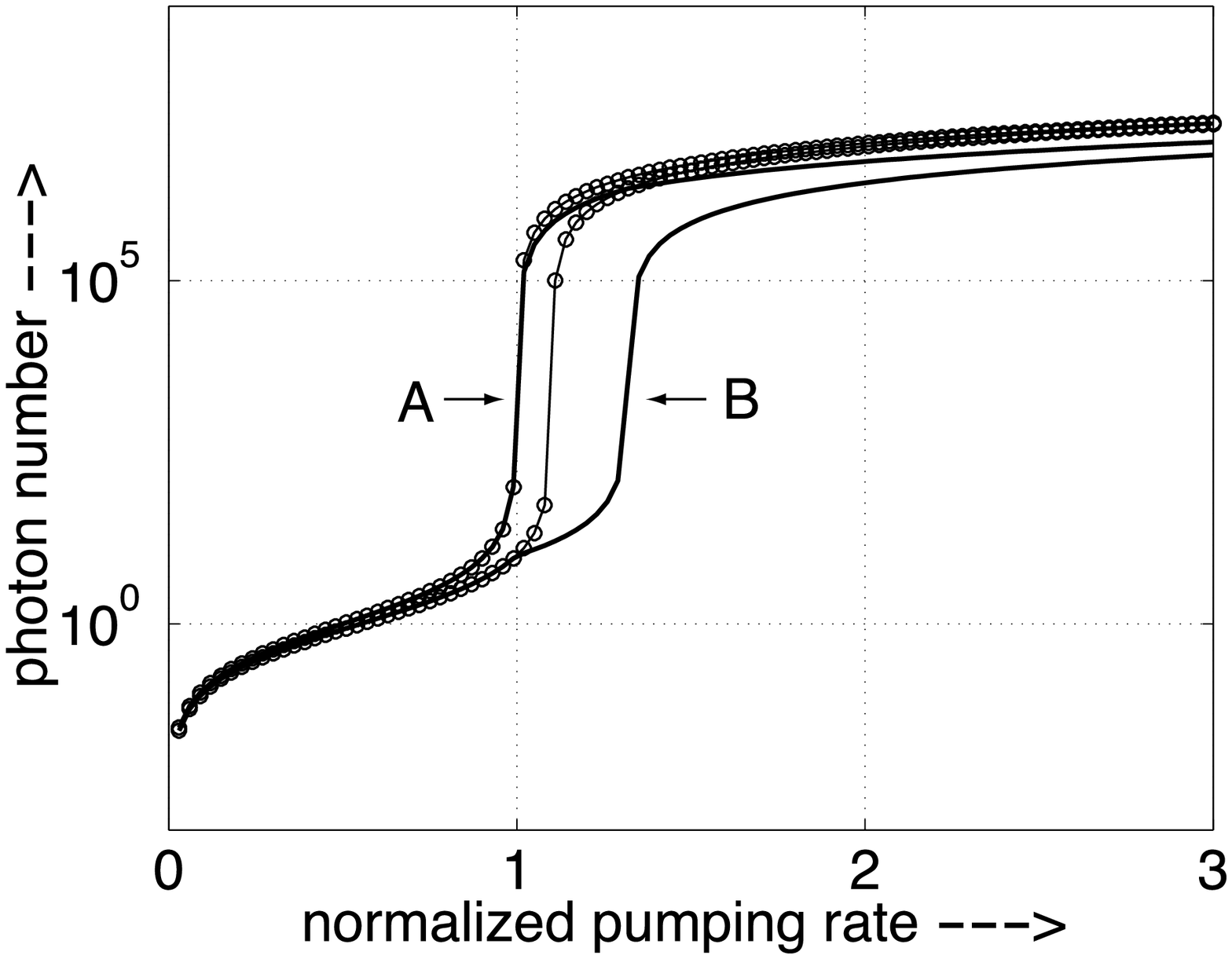}
\hspace{0.35in} 
\includegraphics[width=3in]{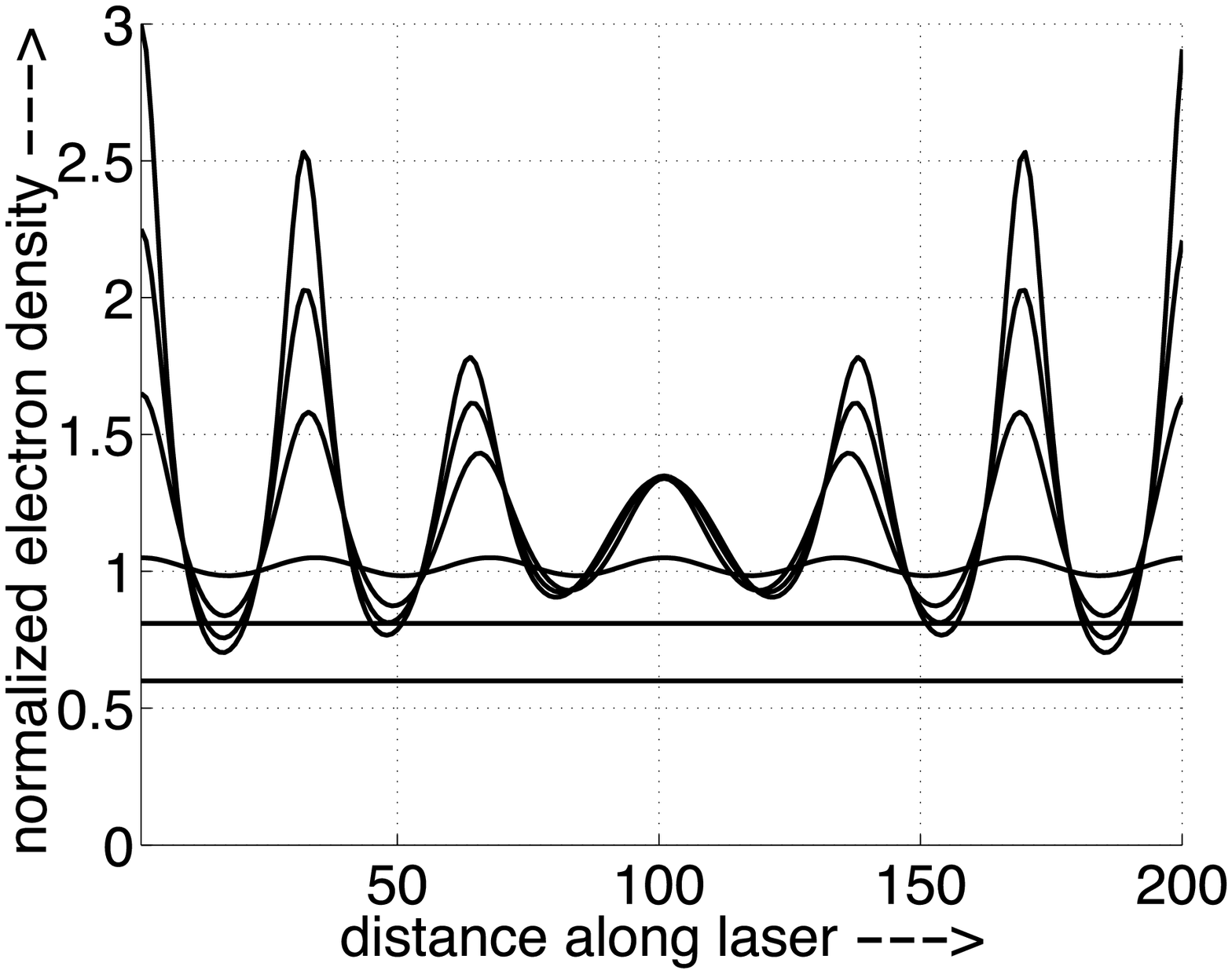}
\caption{(a) Photon numbers and (b) electron density $\rho(x)$ 
for two higher lying longitudinal modes $A=6$ and $B=7$.
The photon numbers versus pumping in (a) depend only weakly on the
number of half wavelengths in the laser cavity. Spatial holes are
burned into the electron density $\rho(x)$ in (b), due both to the
mirrors and the oscillating optical mode intensities.}
\label{semdens}
\end{figure}

Figure~\ref{semdens}(a) shows the photon numbers $n_A$ and $n_B$
versus pumping and for two higher lying longitudinal
modes having $A=6$ and $B=7$. The photon numbers $n_A$ and $n_B$
in Fig.~\ref{semdens}(a) are essentially unchanged from those for
the two lowest cavity modes having $A=1$ and $B=2$ in 
Fig.~\ref{sema1b2}(b). Figure~\ref{semdens} and Fig.~\ref{sema1b2}
use the same parameters, namely $z=1.1$ and $p=10^7$.
The photon numbers $n_A$ and $n_B$ versus pumping
therefore have little (if any)
dependence on the number of half wavelengths in the cavity. 
The weak dependence of Fig.~\ref{sema1b2}(a) on the number of
half wavelengths (where $A-B=1$) is because the fraction
of the gain media where $|u_B|^2 \ge |u_A|^2$ is essentially
independent of the number of half wavelengths in the cavity, 
as can be checked numerically.

Spatial holes are burned into the electron density $\rho(x)$ in 
Fig.~\ref{semdens}(b), especially
when mode $B$ begins lasing. Figure~\ref{semdens}(b) shows
the electron density $\rho(x)$ 
inside the active laser medium for different pumping rates $r$. 
Because the mode intensities have a
node at the mirrors and there is no electron diffusion, the pumping
rates can be read directly from the normalized density axis at the mirrors
(points $l=0$ and $l=200$) in Fig.~\ref{semdens}(b).
The normalized pumping rates in Fig.~\ref{sema1b2}(b) are 
$r=0.6, 0.8, 1.05, 1.65, 2.25,3.0$. 
The electron density $\rho(x)$ is essentially constant for pumping rates
below threshold ($r=0.6, 0.8$) in Fig.~\ref{sema1b2}(b). There is a small
variation in electron density for pumping rates below threshold, which is
invisible on the scale in Fig.~\ref{semdens}(b). Above threshold the variation
in electron density becomes quite pronounced, especially when mode $B$ begins
lasing ($r=1.65, 2.25,3.0$). The growth of electron density $\rho(x)$ as
we move from the center of the gain media towards the mirrors is due to our
neglect of diffusion. Since the optical mode intensites $|u_A|^2$ and 
$|u_B|^2$ have a node at the mirrors, a large spatial hole is also burned into 
the main body of the laser. 
Smaller spatial holes arising from the oscillating optical mode intensities
produce oscillations in the electron density.

\subsection{Fast Diffusion}
\indent

When we include diffusion $(D \ne 0)$, we can no longer solve
Eq.~(\ref{homorho}) directly for the density $\rho(x)$. Instead we
discretize the active lasing medium, taking lattice points $x_l=la$.
Here $a$ is the lattice spacing and $0 \le l \le l_{max}$, with
$L_a = l_{max} a$ the length of the active medium. With this lattice 
Eq.~(\ref{homorho}) reads
\begin{equation}
\frac{d}{dt}
\left[
\begin{array}{c}
\vdots \\
\rho_{l-1} \\
\rho_l \\
\rho_{l+1} \\
\vdots 
\end{array}
\right]
=
\left[
\begin{array}{ccccc}
\ddots & \vdots & \vdots & \vdots & \cdots \\
t & -2t + w_{l-1} & t & 0 & 0 \\
0 & t & -2t + w_{l} & t & 0 \\
0 & 0 & t & -2t + w_{l+1} & t \\
\cdots & \vdots & \vdots & \vdots & \ddots  \\
\end{array}
\right]
\left[
\begin{array}{c}
\vdots \\
\rho_{l-1} \\
\rho_l \\
\rho_{l+1} \\
\vdots  
\end{array}
\right]
+
\left[
\begin{array}{c}
\vdots \\
R_{l-1} \\
R_{l} \\
R_{l+1} \\
\vdots 
\end{array}
\right].
\label{rhodtd}
\end{equation}
Here $t=D/a^2$ is the diffusion rate and 
\begin{equation}
-w_l = \tilde{K}_A V_L n_A |u_A(x_l)|^2 
+ \tilde{K}_B V_L n_B |u_B(x_l)|^2 + \tilde{A}.
\end{equation}
Given an initial guess for the photon numbers $n_A$ and $n_B$, we
can invert Eq.~(\ref{rhodtd}) for the density $\rho(x)$ in steady state.
Taking a hypothetical five point lattice we have 
\begin{equation}
-\left[
\begin{array}{c}
\rho_0 \\
\rho_1 \\
\rho_2 \\
\rho_3 \\
\rho_4 
\end{array}
\right]
=
\left[
\begin{array}{ccccc}
-t + w_0 & t & 0 & 0 & 0 \\ 
t & -2t + w_1 & t & 0 & 0 \\
0 & t & -2t + w_2 & t & 0 \\
0 & 0 & t & -2t + w_3 & t \\
0 & 0 & 0 & t & -t + w_4 \\
\end{array}
\right]^{-1}
\left[
\begin{array}{c}
R_0 \\
R_1 \\
R_2 \\
R_3 \\
R_4
\end{array}
\right].
\label{rhodss}
\end{equation}
We use zero derivitive
boundary conditions to truncate the matrix in Eq.~(\ref{rhodss}). 
After solving Eq.~(\ref{rhodss}) numerically for $\rho_l$, we can substitute
this density back into Eqs.~(\ref{nrat})-(\ref{nrbt}) to generate updated
photon numbers $n_A$ and $n_B$ in steady state. We move from the $m$th to
the $(m+1)$st iteration for the photon numbers by
\begin{equation}
n_A(m+1) = (n_A(m)+1) \int_a \frac{[\tilde{K}_A V_L]}{\gamma_A} |u_A|^2 \rho dV,
\label{dnrat}
\end{equation}
and
\begin{equation}
n_B(m+1)= (n_B(m)+1) \int_a \frac{[\tilde{K}_B V_L]}{\gamma_B} |u_B|^2 \rho dV.
\label{dnrbt}
\end{equation}

\begin{figure}[htb]
\includegraphics[width=3in]{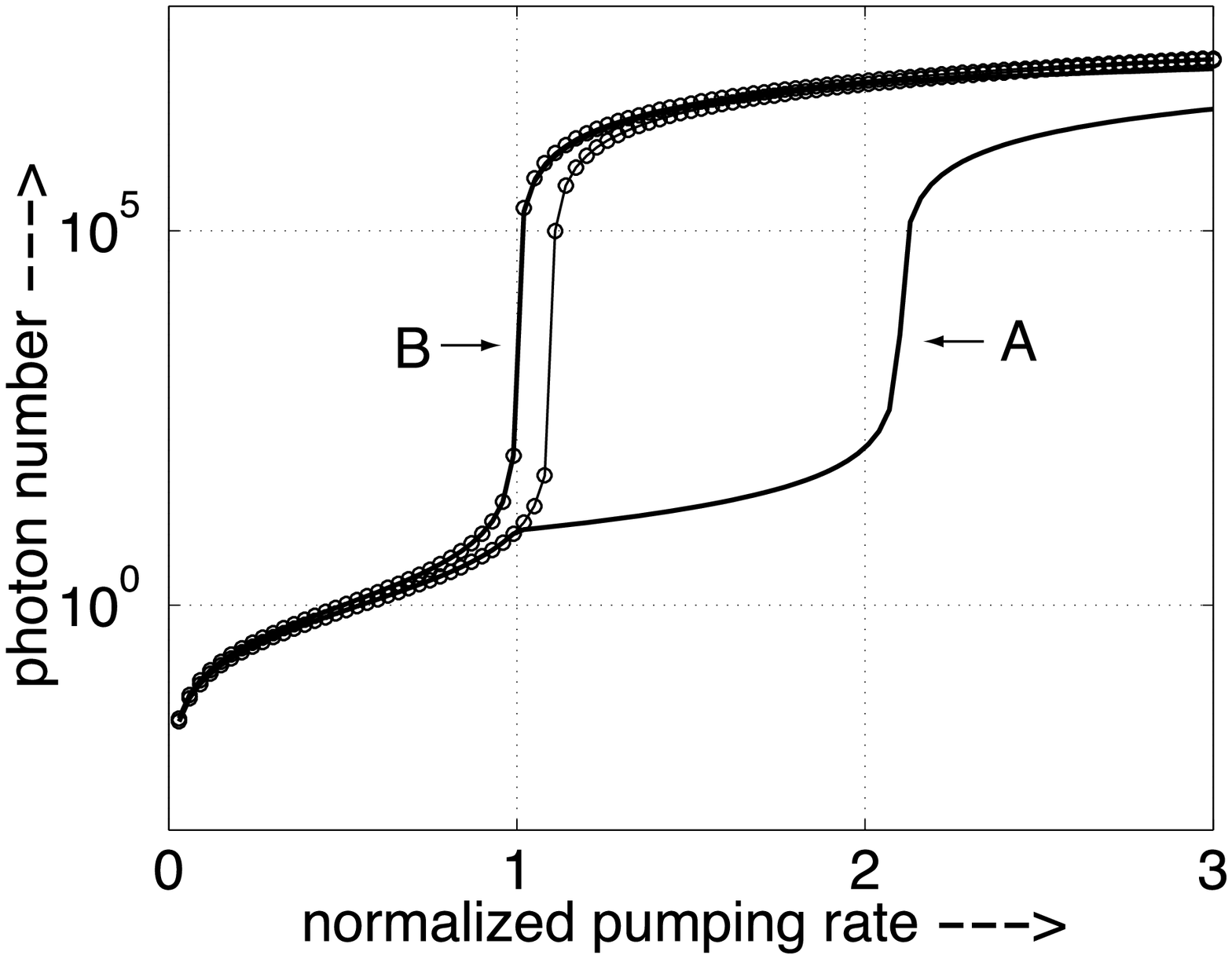}
\hspace{0.35in} 
\includegraphics[width=3in]{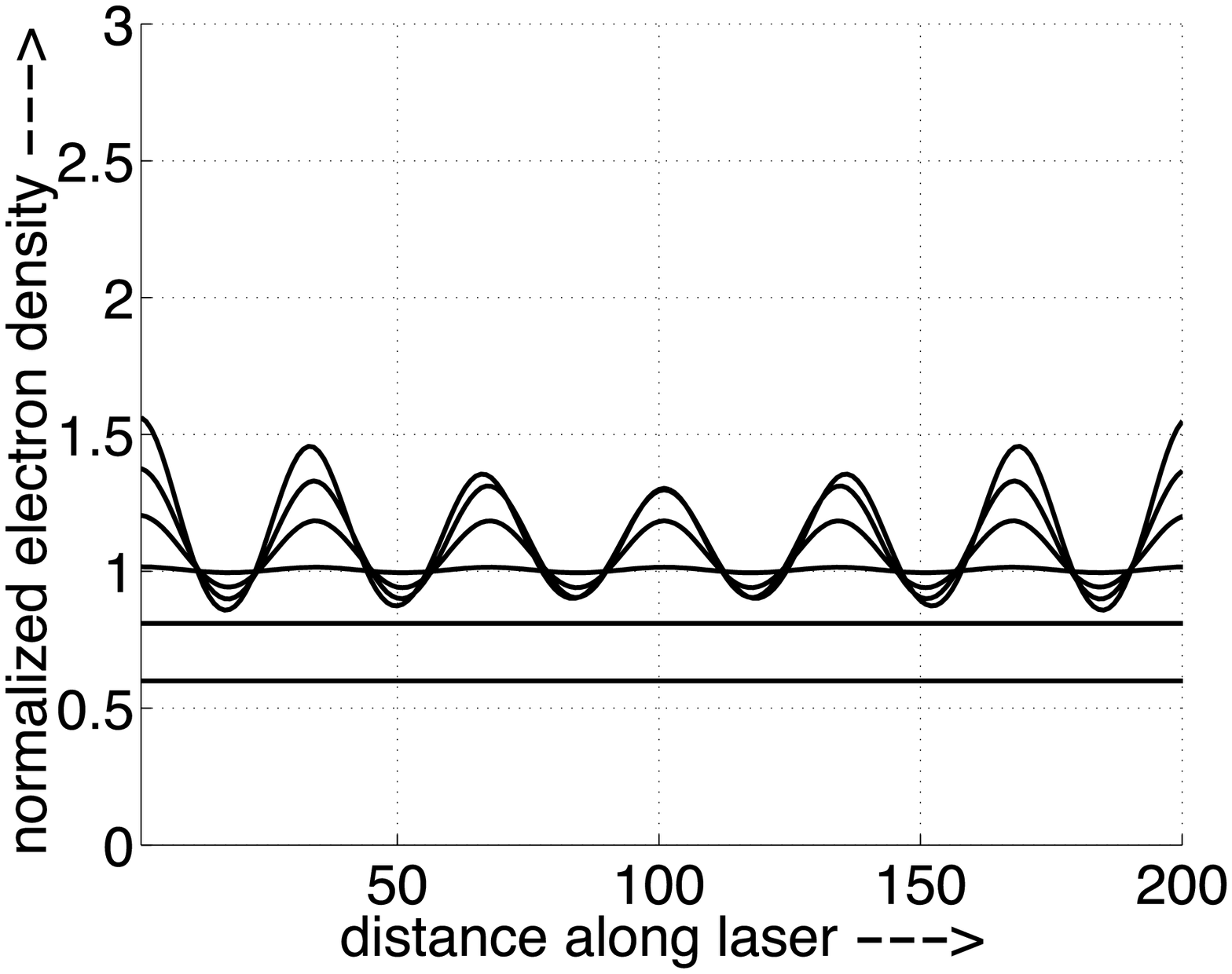}
\caption{(a) Photon numbers and (b) electron density
$\rho(x)$ when the electron diffusion constant
is $D = (9/400) \lambda^2 \tilde{A}$. Adding some electron diffusion
has raised the threshold current for mode $B$ and
reduced spatial hole burning effects due to the mirrors.}
\label{semidiff1}
\end{figure}

\begin{figure}[htb]
\includegraphics[width=3in]{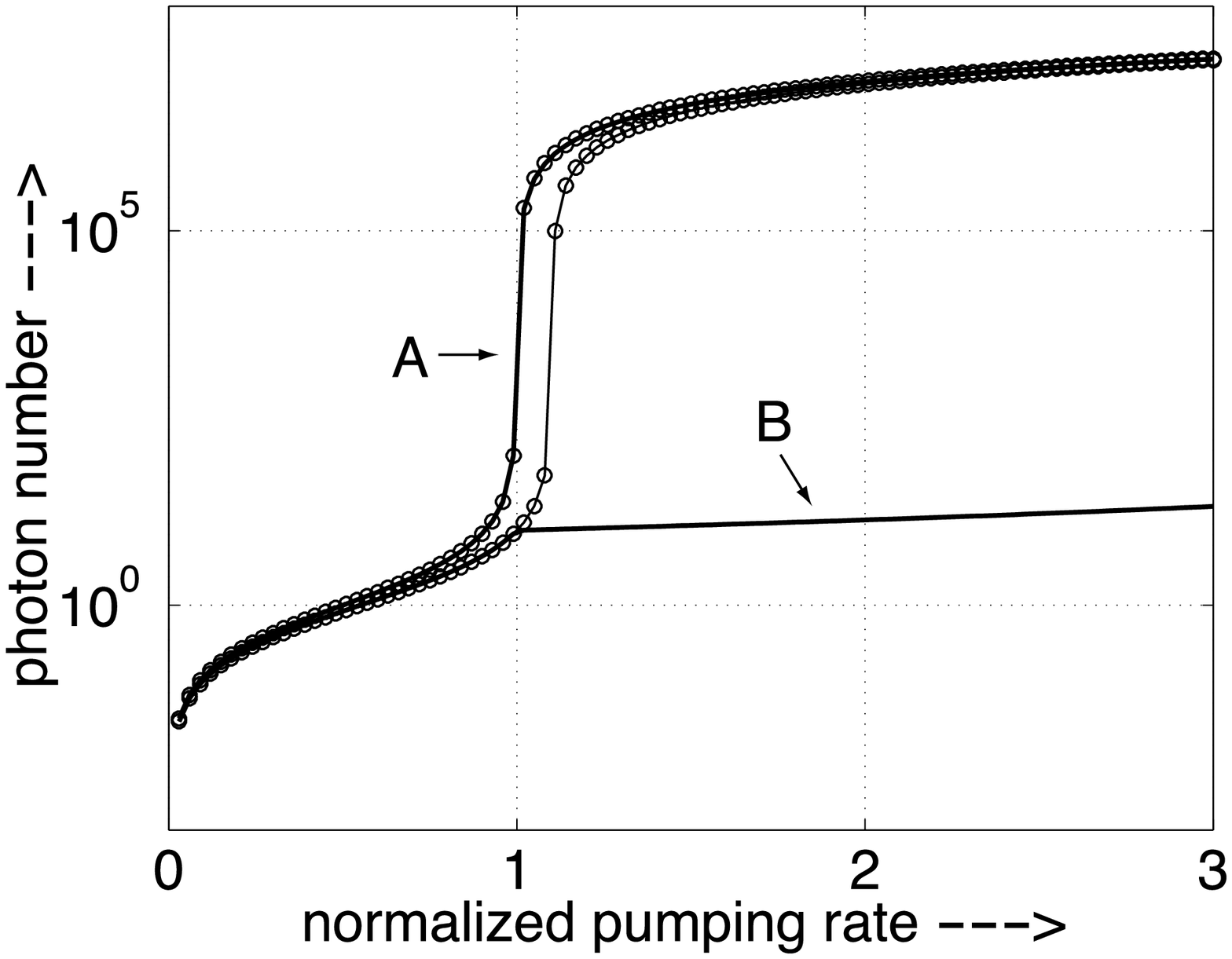}
\hspace{0.35in} 
\includegraphics[width=3in]{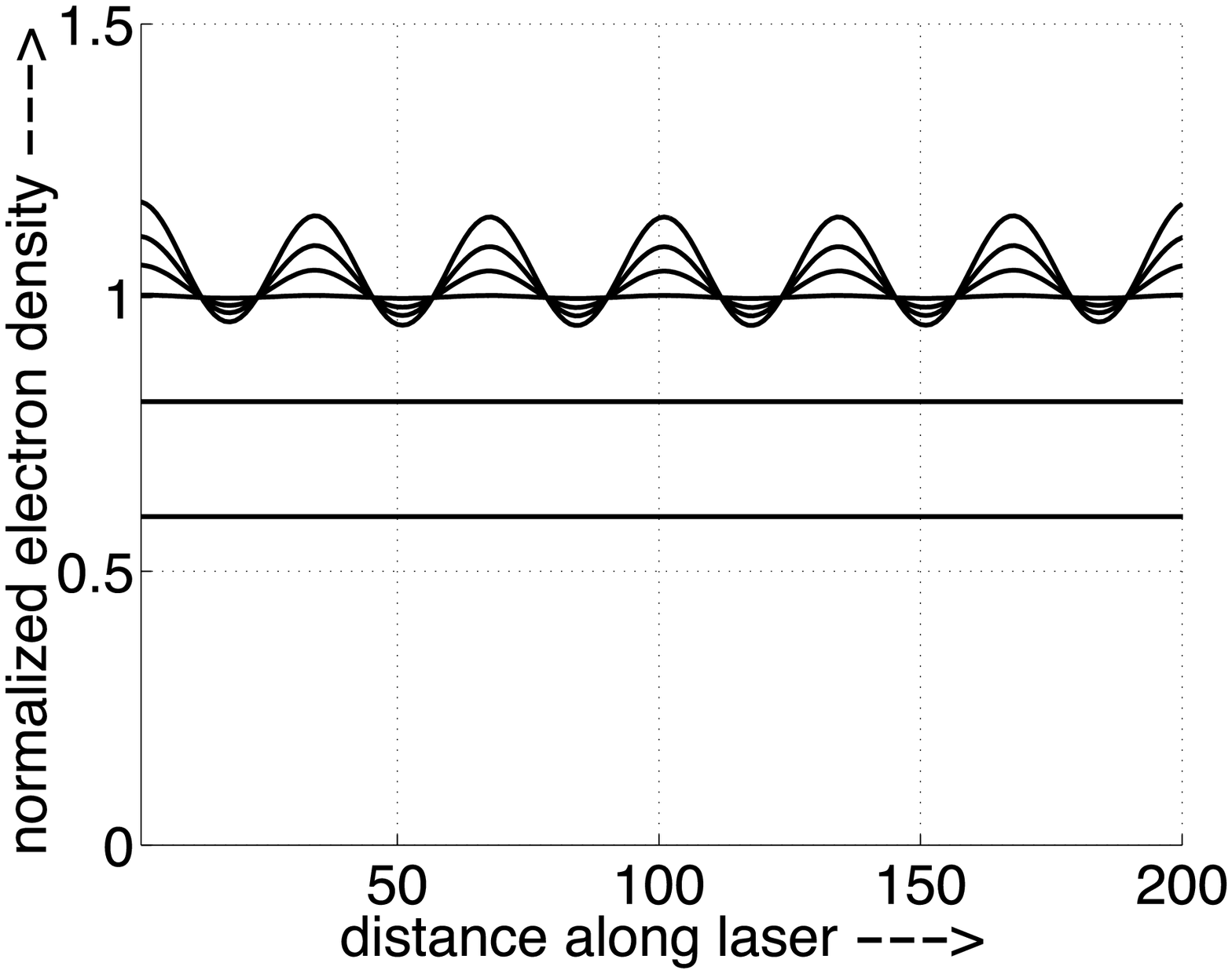}
\caption{(a) Photon numbers and (b) electron density
$\rho(x)$ when the electron diffusion constant
is $D = (45/400) \lambda^2 \tilde{A}$. The diffusion constant is
now large enough that mirror effects are negligible. The electron density
oscillates essentially periodically and the photon number in mode $B$ is
essentially constant above threshold.}
\label{semidiff2}
\end{figure}

Figures \ref{semidiff1} and \ref{semidiff2} show the effects of 
adding electron diffusion to the photon numbers and electron density
in Fig.~\ref{semdens}. In Figs.~\ref{semidiff1}-\ref{semidiff2} 
the number of half wavelengths in the cavity are $A=6$ and $B=7$
with a ratio of optical rate constants $z= \tilde{K}_A/\tilde{K}_B =1.1$,
the same parameters as in Fig.~\ref{semdens}. Figure~\ref{semidiff1} uses
a diffusion constant $D = 100 a^2 \tilde{A}$ with $L = 200 a$ ($t=100p$ with
$l_{max} = 200$), while Fig.~\ref{semidiff2} uses a larger diffusion constant
$D = 500 a^2 \tilde{A}$ with $L = 200 a$ ($t=500p$ with $l_{max} = 200$).
Since the cavity is six half wavelengths long ($L = 6 \lambda /2$), 
we have $D = (9/400) \lambda^2 \tilde{A}$ in Fig.~\ref{semidiff1} and 
$D = (45/400) \lambda^2 \tilde{A}$ in Fig.~\ref{semidiff2}. These values
for the diffusion constant are in good agreement with our order of
magnitude estimate for when diffusion should affect the laser output 
characteristics.

Adding electron diffusion raises the threshold current required for mode $B$
to lase, as can be seen in Figs.~\ref{semidiff1}(a) and \ref{semidiff2}(a). 
The larger the diffusion constant, the greater is the threshold current required
for mode $B$ to begin lasing. Diffusion also reduces the spatial hole
burning effects due to the mirrors, leaving only the smaller spatial holes
due to the difference in the number of half wavelengths in the cavity between
modes $A$ and $B$. Some overall gradients are still visible in the electron
density in Fig.~\ref{semidiff1}(b), while in Fig.~\ref{semidiff2}(b) the
overall average electron density density is essential uniform (mirror effects
are negligible). Fig.~\ref{semidiff2}(b) resembles the picture of electron density
used to illustrate the effects of spatial hole burning in the laser in 
Ref.~\cite{siegman71}.

\clearpage

\section{Conclusions}
\indent

We have generalized the laser rate equations in Refs.~\cite{siegman71,siegman86}
both electron scattering between the different lasing levels to describe spectral
hole burning effects in gas lasers. In order to model spatial hole burning
effects present in semiconductor lasers, and guided by Ref.~\cite{svelto}, 
we then further
generalized the rate equation model to include the effects
of spatially varying optical mode intensities in the laser.

In order for multiple frequencies to lase simultaneously, either the energy
spectrum or spatial variation of the optical gain must be broken up into
many independent (single moded) lasers. Electron equilibration (scattering rate) 
is slow in gas lasers, and this allows the energy spectrum to be broken up
into many independent frequency ranges. An order or magnitude estimate for single
mode laser operation to occur in gas lasers is that the scattering rate
between electrons in the different energy ranges must exceed the spontaneous
emission rate ($s \gg A$). For semiconductor lasers the electron
diffusion is slow, and the gain media can be viewed as many independent
lasers at each point in space. Due to spatial variations in the optical mode
intensities, different lasing modes will be favored at each point in space.
Since the regions where different modes dominate lasing are spatially separated
by about one quarter wavelength,
we need the diffusion constant to exceed $D \gg \lambda^2 A /16$
for single moded operation in 
semiconductor lasers. Numerical simulations given in this paper 
agree with these two order of magnitude
estimates for the transition from single to multiple moded laser operation.

Finally, we can summarize some general (and well known) 
conclusions about single versus multiple moded laser operation.
Firstly, all lasers are single moded for some range of pumping rates
near threshold. The range of pumping rates for single moded operation
is larger for more scattering between electronic states and for faster
electronic diffusion. But a range
of pumping rates for single moded operation nonetheless
exists no matter how weak the
equilibration or how slow the electronic diffusion (unless two degenerate
states are lasing). Secondly, all lasers become multi-moded when pumped
hard enough (unless the gain medium is first destroyed by too high
of a pumping rate).
Finally, bad economic analogies do not describe laser mode competition.
Statements such as 'Laser mode competition is just like life. The rich get
richer and the poor get poorer.' are clearly incorrect. Even as the photon number
in mode $A$ increases, the worst that can happen is that the photon number in
mode $B$ remains constant. Mode $B$ can also begin lasing (become an economic
success) either by electrons scattering from mode $A$ (working in a supporting
industry often created by a competitor) or by specializing its spatial mode
pattern to take advantage of optical gain inaccessible to $A$
(working in another area of the economy to exploit
talents and resources unavailable to a competitor).

\end{document}